\documentclass[a4paper,aps,twoside,brazil,english,prd,amsmath,nofootinbib,
superscriptaddress,showpacs, preprint]{revtex4-1}

\usepackage[T1]{fontenc}
\usepackage[utf8]{inputenc}
\usepackage[brazil,english]{babel}
\usepackage{tensor}
\usepackage[usenames,dvipsnames]{xcolor}
 \definecolor{darkblue}{RGB}{0,0,150}
\usepackage{amssymb}
\usepackage{amsmath}
\usepackage{tensor}
\usepackage{graphicx}
\usepackage{nicefrac}
\usepackage{amsthm}
\usepackage{mathrsfs}

\usepackage[normalem]{ulem}
\usepackage[unicode]{hyperref}
\hypersetup{
  pdfinfo = {
    Author = {Alan , Morgan, Mimoso},
    Title = {Raychaudhuri 2+2}
  },
  colorlinks = true,
  breaklinks = true,
 linkcolor = blue,
 urlcolor = blue,
 citecolor = blue,
}
\usepackage{multirow}
\usepackage{enumerate}

\usepackage[
            linecolor=orange,
            backgroundcolor=red!30,
            colorinlistoftodos]{todonotes}

\newcommand{\ud}{\ensuremath{\mathrm{d}}}
\newcommand{\Lie}{\ensuremath{\mathcal{L}}}

\DeclareMathAlphabet{\mathpzc}{T1}{pzc}{m}{it}

\def\coxa{{\Huge
C\kern-.1667em\lower.5ex\hbox{O}\-X\kern-.1667em\lower.5ex\hbox{A}\@}%
\index{CoXa}
}


\begin{document}

\begin{abstract}
We explore in detail the 2+2 and 1+1+2 formalism in spherically symmetric spacetimes, spanning from deducing the dynamical equations to relating them to the well-known generalised Painlevé-Gullstrand  (GPG) coordinate system. The evolution equations are the 
Raychaudhuri equations for null rays, including those also known as cross-focusing equations whose derivation, to the best of our knowledge, we present for the first time. We physically interpret the scalars that arise in this scenario, namely the flow 2-expansion $\Theta_{n}$, the flow acceleration $\mathcal{A}$, and  the radial extrinsic curvature $\mathcal{B}$. We derive a coordinate independent formula for the redshift which shows that $\mathcal{B}$ is the sole source for the redshift in spherically symmetric spacetimes. We also establish the correspondence between the 1+1+2 scalars and the 1+3 splitting scalars, expansion and shear. We further make a comparison with the Newman-Penrose formalism, in order to clarify the context where each formalism is more useful, and finally, we extend our results to planar and hyperbolic symmetric warped spacetimes as well, in particular, the relationship between $\mathcal{B}$ and the redshift. 
\end{abstract}
\title{New insights on null and timelike warped symmetric spacetime splittings
}
\author{Alan Maciel$^\star$}
\address{Centro de Matem\'atica, Computa\c c\~ao e Cogni\c c\~ao, Universidade Federal do ABC,\\
 Avenida dos Estados 5001, CEP 09210-580, Santo Andr\'e, S\~ao Paulo, Brazil}

\author{Morgan~Le~Delliou$^{*,\dagger}$}
\address{-Institute of Theoretical Physics \& Research Center of Gravitation, Lanzhou University, Lanzhou 730000, China\\
	-Key Laboratory of Quantum Theory and Applications of MoE, Lanzhou University, Lanzhou 730000, China\\
	-Lanzhou Center for Theoretical Physics \& Key Laboratory of Theoretical Physics of Gansu Province, Lanzhou University, Lanzhou 730000, China} 	
	\address{Instituto de Astrof\'isica e Ci\^encias do Espa\c co, Universidade de Lisboa,
	Faculdade de Ci\^encias, Ed.~C8, Campo Grande, 1769-016 Lisboa, Portugal}
 \address{Universit\'e de Paris, APC-Astroparticule et Cosmologie (UMR-CNRS 7164), 
 F-75006 Paris, France.}
 
\author{Jos\'e P. Mimoso$^\ddagger$}
\address{Instituto de Astrof\'isica e Ci\^encias do Espa\c co, Universidade de Lisboa,
	Faculdade de Ci\^encias, Ed.~C8, Campo Grande, 1769-016 Lisboa, Portugal}
\address{Departamento de F\'{i}sica  \\ 
Faculdade de Ci\^{e}ncias da Universidade de Lisboa, \\ 
Campo Grande, Ed. C8 1749-016 Lisboa,
Portugal\\
E-mail: $^\ddagger$jpmimoso@fc.ul.pt\\
\quad $\star$alan.silva@ufabc.edu.br\\
\quad $\dagger$delliou@lzu.edu.cn, Morgan.LeDelliou.ift@gmail.com
}
\maketitle

\section{Introduction}
One of the fundamental pillars of the theory of General Relativity (GR) is 
the Principle of Covariance~\cite{Penrose:2005bg,dInverno:1992gxs}. It establishes that the laws of Physics ought to be the same, independently of the choice of coordinate system or observers. By the same token it enables one to adopt choices of coordinates that best suit the comprehension of the questions under investigation without contaminating the results. 

Among the various systems of coordinates that can be found in the 
literature~\cite{weinberg}, the consideration of two congruences of null vector fields has been devised to address problems where we focus on the structure of light cones.  It was in due time  perceived that null geodesics determine the behaviour of the latter, and hence define the causal domains. This  was manifest  in connection with the characterization of black holes, where,  in particular,  null coordinates play an essential role in 
the compactification of spacetime in the remarkable Penrose diagrams~\cite{Kopczynski-Trautman-book,hawking}, and in recent results regarding both the thermodynamics of black holes, and more general cosmological settings~\cite{Hayward:1993mw,Bak:1999hd,Cai:2006rs,Hayward:1997jp,Binetruy:2014ela}.
The consideration of two congruences of null vectors forming a tetrad together  with   two complex vectors  also underlies the so-called Newman-Penrose (NP) formalism~\cite{Sachs:1964zza,Newman:1961qr,Chandrasekhar:1980gb}, that is particularly adequate to establish 
the Petrov classification of the Weyl tensor, and for the investigation of gravitational radiation. However the latter formalism is technically involved and difficult to decode. 
The dual-null formalism (alternatively dubbed double-null) has also been adopted in studies of the initial value problem~\cite{dInverno:1980kaa,Brady:1995na}, as well as in connection to the use of the so-called geodesic light-cone (GLC) coordinates ~\cite{Fleury:2016htl,Nugier:2016ebh}  applied to static black holes. The numerical applications of the double-null-coordinates to a number of gravitational problems  was also recently the object of a comprehensive review~\cite{Nakonieczna:2018tih}.

In  previous  work we have utilised  the dual null formalism \cite{Maciel:2015vva}  to extend both the Birkhoff theorem \cite{Maciel:2018tnc} and the Tolman-Oppenheimer-Volkov (TOV) \cite{Maciel:2019grc} equation to more general geometric frameworks, such as planar or cylindrically symmetric space times, as well as hyperbolically symmetric spacetimes.

In addition, when matter dominates the physics of the spacetime, it is useful to unfold the 2+2 formalism into an 1+1+2 formalism that singles out the matter flow.  The 1+1+2 formalism has been used in order to study metric perturbations in symmetric spacetimes \cite{Clarkson:2002jz, Clarkson:2007yp}, variations of the Birkhoff Theorem \cite{Goswami:2011ft,Goswami:2012jf,Ellis:2013dla}, optical and light beam propagation in spacetimes~\cite{Glod:2020adq}, as well as a way to define quasilocal universal horizons \cite{Maciel:2015ypv} in modified gravity theories. Our objective in this work is to connect the characterizations of spacetime dynamics in both frameworks in a clear way, 
to extend the spherically symmetric approach to other symmetries, 
as well as to show we can obtain the usual metric, namely its generalised Painlevé-Gullstrand (GPG) description ~\cite{{Painleve1921,gullstrand-1922},Gautreau:1984PhRvD186,LaskyLun06b,LaskyLun07,Faraoni:2020ehi} 
and coordinate based description, from the matter flow scalars, in a similar manner as done in generalised Lemaître-Tolman-Bondi (gLTB) coordinates \cite{Sussman:2008wx}. We also aim to connect some of these scalars to physically meaningful observables, such as the redshift \cite{Ellis:1971pg, Codur:2021wrt} and to point out the better fit of these formalisms compared with the NP approach \cite{Chandrasekhar:1980gb} to our questions. Finally, we 
present in the appendix the detailed derivations of the evolution and constraint equations in 2+2 formalism, for spherically symmetric spacetimes, first presented in \cite{Hayward:1993mw,hayward-1993}, as its exposition seem to evade the literature.

In the present work, we revisit the 2-null formalism in a more comprehensive way in order to fully display its structure, and aiming at to render its advantages more apparent in the case of warped spaces with symmetric foliations, namely spherical symmetry. We naturally obtain its relation with the 1+1+2 splitting for the same classes of spacetimes, which shares many of the virtues of the 2+2 formalism, while highlighting the existence of a timelike preferred flow. Such clear relation should allow easy translation between lightcones and matter flow problems, enabling causal conclusions for fluid problems as well as a way to obtain an observer centered perspective in optical problems. We also connect those formalisms with the GPG system of coordinates in order to retrieve the usual description of those spacetimes, and to provide us with a physical understanding of the flow related quantities. Although we start focusing on spherically symmetric spacetimes, we subsequently  generalize our methods to cases of spaces with hyperbolic or planar codimension-2 foliations~\cite{Maciel:2018tnc,Maciel:2019grc,herrera-2010a,Herrera:2021wpe,Carot:1999zm}.

The paper is organised as follows: Sec.~\ref{sec:Dynamics} discusses the dynamics of gravity from the 2+2 description to the GPG coordinates.
In Sec.~\ref{sec:basics}, we describe briefly the main assumptions of the 2+2 formalism, followed by the most relevant scalar quantities that arise from the 2+2 setup in Sec.~\ref{sec:flow}. We then proceed to unfold the dual null basis into a timelike and spacelike preferred basis, obtaining the equivalent flow scalars and equations for the 1+1+2 formalism. We relate our quantities with the GPG coordinates, by retrieving the GPG line element from the flow scalars in Sec.~\ref{sec:GPG}. 

Those results lead us to propose physically meaningful interpretations in Sec.~\ref{sec:PhysInterp}. We 
give the physical interpretation of the Raychaudhuri-like evolution and constraint equations in Sec.~\ref{Sec:Interpret}. In Sec.~\ref{sec:flow2}, we also relate the scalars of the 1+1+2 with those of the most commonly used 1+3 formalism. In Sec.~\ref{sec:redshift} we derive the relationship between the redshift and the flow scalars. 

Finally, generalisations and comparisons of our results are proposed in Sec.~\ref{sec:extensionsComp}, before concluding. We first 
obtain the analogous results for planar and hyperbolic symmetric spacetimes, which are readily obtained in 1+1+2 formalism, in Sec.~\ref{sec:hyperbolic}. Finally, we compare our formalism with the Newman-Penrose formalism, in order to discuss the context where each of them is more useful, in Sec.~\ref{sec:advantages}. In the appendix we show the detailed derivations of the evolution and constraint equations in 2+2 formalism, for spherically symmetric spacetimes.

\section{Dynamics from 2+2 to 1+1+2 and GPG}\label{sec:Dynamics}
\subsection{Basics of the dual null formalism} \label{sec:basics}

We consider spacetimes whose metric may be written in the following form:
\begin{gather}
g_{ab} = n_{ab} + s_{ab}\,, \label{eq:generalmetric}
\end{gather}
\noindent
where $s_{ab} = r^2 \gamma_{ab}$, with $\gamma_{ab}$ the metric of 2-sphere and $s^{ab} \partial_a r = 0$, that is, $r$ constant along the tangent directions to the spheres of symmetry. Those spacetimes admit a codimension-2 
foliation where the leaves correspond to spherical surfaces.

In this dual null formalism  we define two basis vectors $(k^a , l^a)$ that span the subspace orthogonal to the spherical surfaces, formally
\begin{gather}
k^a s_{ab} = l^a s_{ab} = 0\,.
\end{gather}
 Note that unlike the expansion of timelike flow, where if the flow is given by a timelike vector field $u^a$, we can normalize it by imposing $u^{a}u_{a} = -1$, in the null case we have no natural normalization, 
 so that the value of the expansion \emph{is dependent on our choice of a null vector field tangent to the flow}. 
Instead, in the latter case we can choose a kind of normalization fixing the product $k^a l_a  <0$, 
which we are so far keeping arbitrary, albeit constant and negative, 
in order to make the basis future-directed. 

In terms of $k^a$ and $l^a$, the metric \eqref{eq:generalmetric} can be written as
\begin{equation}
 g_{ab} =   \frac{1}{k^c l_c}2k_{(a} l_{b)} + s_{ab}\,,  \label{kruskalmetric}
\end{equation}
\noindent
where, by definition $x_{a}= g_{a b}x^{b}$ and  $x_{(ab)} =(1/2)( x_{ab}+x_{ba})$. Note that the induced metric is independent of the particular choice of $k^a$ and $l^a$, provided they are null.


\subsection{Flow scalars} \label{sec:flow}
The expansion associated with the spacelike two-surfaces along $k^a$ is given by 
\begin{equation}
 \Theta_{(k)} = \frac{1}{2}s^{ab}\Lie_{k}s_{a b}= s^{a b} \nabla_{a} k_{b}  \,,\label{expansion}
\end{equation}
where We denote $\mathcal{L}_{v}$ the Lie derivative along $v^{a}$. 

We conveniently decompose the projected gradient of the null flow -- 
say of $k^a$ -- in a scalar, a symmetric tensor, and an anti-symmetric tensor on the two dimensional leaves:
\begin{equation}
 s\indices{^{a} _{c}} s\indices{^{b} _{d}} \nabla_{a} k_{b} = \frac{\Theta_{(k)}}{2} s_{c d} + \sigma_{c d} + \omega_{c d}\,, \label{decompose_nabla_k}
\end{equation}
\noindent
where the shear $\sigma_{ab}$ is trace-free and symmetric,  and  where the vorticity $\omega_{ab}$ is also trace-free, but anti-symmetric. Remark that, as it happens with the expansions, in the codimension-2
foliation case we have two shear and vorticity tensors, one corresponding to evolution in each of the null directions. In other words, there will be a corresponding decomposition to (\ref{decompose_nabla_k}) for $l_a$.

In the spherically symmetric case, the only non-vanishing component of the gradient of null curves is the expansion as the other terms would violate spherical symmetry.

We can redefine our null basis, by making 
\begin{gather}
    k'^a = \lambda( u_+, u_-) k^a\,\quad \text{and} \quad l'^a = \lambda^{-1} (u_+, u_-) l^a\,, \label{eq:boost}
\end{gather}
\noindent where $u_\pm$ are coordinates along the null directions of $k^a$ and $l^a$, while maintaining our normalization. Then the expansion transforms as
\begin{gather}
 \Theta_{(k')} = \lambda \Theta_{(k)}\,,
\end{gather}
\noindent which will allow us to transform the equations involving flow scalars 
in accordance to a transformation between pairs of dual null basis. The transformation~\eqref{eq:boost} is a local boost transformation with rapidity given by $\ln \lambda$. This freedom is crucial in order to make the unfolding of the 2+2 formalism to the 1+1+2 formalism, since we need to have the liberty to chose the correct null basis that will be related to the matter flow $n^a$ by the use of those local boosts.

The aim is to characterize the dynamics of the spacetime in terms of the relationship of the flow scalars with the matter, instead of the metric, as usual in GR. For this, we combine the well known Raychaudhuri equation for null congruences, which provides de evolution of the null expansions along the null flow, along with
what we call the cross-focusing equations, that gives the behaviour of one null expansion along the other null flow. They are given by
\begin{subequations}
\begin{align}
 \Lie_{k}\Theta_{(k)} &=  - \frac{\Theta_{(k)}^2}{2} +  \nu_k \Theta_{(k)} - 8\pi T_{cd}k^c k^d\,, \label{eq:Raychaudhurikk} \\
 \Lie_l \Theta_{(k)} &= -\Theta_{(k)} \Theta_{(l)} - \nu_l \Theta_{(k)} - \frac{k^c l_c}{r^2} + 8 \pi T_{ab}k^al^b\,,\label{eq:cross}
\end{align} \label{eqs:Raynull}
\end{subequations}
\noindent
where $T_{ab}$ is the energy-momentum tensor (EMT) of the sources, together with two more equations 
obtained by exchanging the roles of $k$ and $l$ in \eqref{eqs:Raynull}. A detailed derivation of those equations is given in Appendix~(\ref{AppRay} and \ref{AppCross}).

We notice that $T_{cd}\,k^c k^d\equiv \mu$ is the complete projection of the energy-momentum tensor of the sources along the null vector $k^a$. It can be seen as a null dust energy density, interpreting $T_{ab}$ as if it were representing null dust along the $l$ direction \cite[Sec.~7.4]{Faraoni:2018fil,EllisMaartensMacCallum2012}. Given the equations~\eqref{n&e}  below, relating the null vectors $k^a$ and  $l^a$ to both the 4-velocity of the matter flow and the spatial vector defining the direction of local rotational symmetry, it is possible to express the quantity $\mu$ under consideration as a combination of the thermodynamical quantities devised by an observer comoving with the matter fluid. The same applies to $T_{ab}\,k^al^b \equiv Q$ that corresponds to the flux of the vector $T_{ab}\, k^a$ accross the wave-surfaces orthogonal to $l^a$, as well as equivalently to the flow of the vector $T_{ab}\,l^a$ accross the wave-surfaces orthogonal to $k^a$. The null dust interpretation would yield 0 by definition, so it corresponds to a more complex picture.

Further notice that, as shown in the continuation of the paper, the sum and differences of Eqs.~\eqref{eqs:Raynull} involve respectively the radial acceleration and the redshift generator scalars.

Incidentally, we recognise that $\mu$ is associated with the  so-called null energy condition~\cite{hawking} which requires  $\mu\ge 0$.

\subsection{Equations in the 1+1+2 formalism} \label{subsec_1+1+2}

Although the dual null formalism, or 2+2, is very convenient in order to study the geometric structure of the spacetime, specially in issues related with strong gravity near dynamical black holes, when there is a matter flow dominating the physics it is usually more relevant to single out the matter flow $n^a$. This leads naturally to the well known $1+3$ formalism. However, in order to keep the advantages of both $2+2$ and $1+3$ formalism, for spacetimes that exhibit spherical symmetry, we apply the $1+1+2$ formalism.
We define, instead of a dual null basis for the subspace orthogonal to the spherical leaves, a basis consisting of a timelike and a spacelike vector, in the following way
\begin{gather}
n^a =-\frac{1}{2 k^c l_c} \left( k^a + l^a \right)\,, \nonumber\\
e^a =-\frac{1}{2 k^c l_c} \left(k^a - l^a \right) \, . \label{n&e}
\end{gather}
\noindent such that
\begin{gather}
    n^a n_a = -1 \,, \quad e^a e_a = 1 \, \quad \text{and} \nonumber\\
    g_{ab} = - n_a n_b + e_a e_b + s_{ab}\,. \label{eq:metric112}
\end{gather}

For the sake of simplicity, from now on we choose $k^c l_c  = -1/2$ as the normalization of the dual null basis. Therefore, we find for the Lie Derivatives with respect to $n^a$ and $e^a$:
\begin{gather}
    \Lie_n = \Lie_k + \Lie_l \,, \nonumber\\
    \Lie_e = \Lie_k - \Lie_l\,. \label{eq:Lie112}
\end{gather}

Although the expansion in relation to a two dimensional spatial orthogonal subspace is natural for null congruences, here we extend  this concept to any vector orthogonal to $s_{ab}$. For any vector $v^a$ generated by the basis $(n^a, e^a)$ this is achieved by defining  
\begin{gather}
\Theta_{(v)} = \frac{1}{2} s^{ab} \Lie_v s_{ab} = \frac{2}{r} v^{a} \nabla_a r\, .
\end{gather}
We call the latter as  2-expansions in order to distinguish them from the usual expansions in 1+3 formalism.

We can also understand the 2-expansions as projections of the mean curvature vector, denoted $\Theta^c$, along the given direction. Since the second fundamental form, also called shape tensor, is 
\begin{gather}
\mathcal{K}\indices{^c_a _b} =  \left( s\indices{_a^d} s\indices{_b^f} \nabla_d N\indices{_f^c} \right)
=  -\left(  k^c\Lie_l s_{ab} +  l^c \Lie_k s_{ab}\right) \,,
\end{gather}
\noindent where  $N\indices{_a^b} = -2\left(k_a l^b + l_ak^b\right)$ is the projector onto the subspace orthogonal to $s_{ab}$, the mean curvature vector, $\Theta^c$, is given by the trace of $\mathcal{K}\indices{^c_a_b}$ thus yielding 
\begin{gather}
\Theta^c = -2\Theta_{(k)}l^c - 2\Theta_{(l)}k^c\,.
\end{gather}

So, we can now equivalently define the 2-expansions $\Theta_{(v)}$ as the projections of $\Theta^c$ along $v^a$:
\begin{gather}
\Theta_{(v)} \equiv v_c \Theta^c\,.
\end{gather}
In particular, the 2-expansions along $n^a$ and $e^a$ are given by
\begin{gather}
    \Theta_{(n)} = \Theta_{(k)} + \Theta_{(l)}\,, \nonumber\\
    \Theta_{(e)} = \Theta_{(k)} -\Theta_{(l)}\,.
\end{gather}

We aim at computing the Raychaudhuri-like equations along the directions $n^a$ and $e^a$, as well as the mixed equations. By using Eqs. \eqref{eqs:Raynull}, we obtain the relations

\begin{subequations}
\begin{align}
\Lie_n \Theta_{(n)} = \left( \Lie_{k} + \Lie_{l} \right) \left( \Theta_{(k)} + \Theta_{(l)}\right)\, ,  \\
\Lie_e \Theta_{(e)} =\left( \Lie_{k} - \Lie_{l} \right) \left( \Theta_{(k)}- \Theta_{(l)} \right) \,,\\
\Lie_n \Theta_{(e)} = \left( \Lie_{k} + \Lie_{l} \right) \left( \Theta_{(k)} - \Theta_{(l)} \right)\,,\\
\Lie_e \Theta_{(v)} = \left( \Lie_{k} - \Lie_{l} \right) \left( \Theta_{(k)} + \Theta_{(l)} \right)\,,
\end{align}
\end{subequations}
\noindent
which lead to
\begin{subequations} \label{eqs:RayNE}
\begin{align}
   \Lie_n \Theta_{(n)}& = - \frac{3\Theta_{(n)}^2}{4} + \frac{ \Theta_{(e)}^2}{4} - \frac{1}{r^2}+ \mathcal{A}\Theta_{(e)} - 8 \pi T_{ab}e^a e^b \,, \label{eq:RayNE-a}\\
 \Lie_e \Theta_{(e)} &= -\frac{3\Theta_{(e)}^2}{4} + \frac{\Theta_{(n)}^2}{4} + \mathcal{B} \Theta_{(n)} + \frac{1}{r^2} - 8\pi T_{ab}n^a n^b\,,\label{eq:RayNE-b}\\
 \Lie_n \Theta_{(e)} &=  - \frac{\Theta_{(n)} \Theta_{(e)}}{2} + \mathcal{A} \Theta_{(n)}  - 8 \pi T_{ab}n^a e^b \,,\label{eq:RayNE-c}\\
 \Lie_e \Theta_{(n)} &= \Lie_n \Theta_e + \mathcal{B}\Theta_{(e)} -\mathcal{A}\Theta_{(n)}\,,\label{eq:RayNE-d}
\end{align}
\end{subequations}
\noindent
where we have defined
\begin{gather}
    \mathcal{A} = n^a e^b \nabla_a n_b= -2 (k^c l_c) \left(\nu_k - \nu_l \right)= \nu_k - \nu_l \,, \nonumber\\
    \mathcal{B} =e^a e^b \nabla_a n_b = -2(k^c l_c) \left( \nu_k + \nu_l \right) = \nu_k + \nu_l\,.\label{eq:AandB}
\end{gather}

We recognize $\mathcal{A}$ as the flow scalar acceleration, and we also  realise that $\mathcal{B}$ can also be defined as $\mathcal{B}= \mathcal{K}_{ab}e^a e^b$, where $\mathcal{K}_{ab}$ is the extrinsic curvature of the hypersurface orthogonal to the flow\footnote{The extrinsic curvature is equivalent to the shape tensor in codimension-one foliations, the shape tensor in this case is given by $\mathcal{K}\indices{^c _a _b} = n^c\mathcal{K}_{ab}$.} and, as it will be shown in Sec.~\ref{sec:redshift}, it is also the scalar that acts as source for the redshift.


\subsection{From 1+1+2 to the GPG metric}\label{sec:GPG}

The 1+1+2 formalism presents many advantages in order to analyse the dynamical equations of the spacetime and predict its behavior. By using only coordinate independent quantities that are adapted to the symmetries of the spacetime, it  not only highlights the matter flow, as in  the 1+3 threading, but also puts in evidence the spatial direction orthogonal to the symmetric surfaces. However, for practical purposes 
it is important to relate this approach 
to well known coordinates system for spherically symmetric spacetimes, such as the generalized Painlevé-Gullstrand (GPG) coordinates. In order to achieve this goal, we recall the definition of the Misner-Sharp mass-energy $M_{ms}$ \citep{MisnerSharp}:
\begin{gather}
g^{ab} \partial_a r \partial_b r = 1 - \frac{2 M_{ms}}{r}\,, \label{M_ms-def1}
\end{gather}
which, together with Eq.~\eqref{eq:metric112}, gives us
\begin{gather}
 - (n^a \partial_a r)^2 + (e^b \partial_b r)^2 = 1 - \frac{2 M_{ms}}{r} \,, \label{eq:M_ms_develop}
\end{gather}
which we may rearrange as
\begin{gather}
e^b \partial_b r = \pm \sqrt{ 1 + (n^a \partial_a r)^2 - \frac{2 M_{ms}}{r}}\,. \label{eq:epartialr}
\end{gather}

Interpreting the term $n^a \partial_a r$ as the fluid radial velocity, we are 
then 
motivated to define
\begin{gather}
E = (n^a \partial_a r)^2 -  \frac{2 M_{ms}}{r}\,, \label{eq:Edefinition}
\end{gather}
\noindent which corresponds to twice the Newtonian mechanic energy per unit mass\footnote{In some sources $E$ is defined as the Newtonian energy per unity mass, but here we prefer to avoid a factor 2 appearing in the GPG metric.}.

Using the scalar function $E$, Eq.~\eqref{eq:epartialr} becomes
\begin{gather}
e^b \partial_b r = \pm \sqrt{1 + E}\,. \label{eq:epartialrE}
\end{gather}

The GPG coordinates correspond to using a timelike coordinate $t$ chosen along the flow, 
together with the areal radius as the spacelike coordinate orthogonal to the spheres of symmetry. Therefore, translating from the abstract index notation to a coordinate notation, we have
\begin{gather}
n_a \rightarrow - \alpha(t,r) \ud t \,,\label{eq:ntranslation}
\end{gather}
\noindent
where we chose the minus sign in order to guarantee that $t$ is future directed with positive $\alpha(t,r)$, which corresponds to the shift function 
of the Arnowitt-Deser-Misner  (ADM) decomposition \cite{Arnowitt:1959ah}. Since, by construction, $n_a e^a = 0$, we realise that $e^a$ has the form $(0, e^r, 0, 0)$. From 
Eq.~\eqref{eq:epartialrE}, we conclude that
\begin{gather}
e^r = \sqrt{1+E}  \Rightarrow  e^a \rightarrow \sqrt{1+E(t,r)} \partial_r\,, \label{eq:etranslationup}
\end{gather}
\noindent
where we chose the plus sign in order to make $e^a$ directed outwards, that is, towards increasing $r$.

In order to translate the full metric to coordinate notation we have to find the form of $e_a$. Since $e^a$ is a unity vector, applying $e^a e_a = 1$ gives us $e_a = (e_t, \frac{1}{\sqrt{1+E}}, 0, 0)$, with $e_t$ unconstrained. Therefore, defining a function $\beta(t,r) =- e_t \sqrt{1+E} $, we write:
\begin{gather}
e_a \rightarrow \frac{1}{\sqrt{1+E(t,r)}}\left(- \beta(t,r)\ud t + \ud r \right)\,. \label{eq:etranslation}
\end{gather}

Plugging Eqs.~\eqref{eq:ntranslation} and \eqref{eq:etranslation} into Eq.~\eqref{eq:metric112} and making the final translation
\begin{gather}
s_{ab} \rightarrow r^2 (\ud \theta^2 + \sin^2 \theta^2 \ud \varphi^2) = r^2 \ud \Omega^2\,,
\end{gather} 
\noindent
we obtain the well known GPG line element \cite{Lasky:2006zz,adler-2005,LaskyLun06b,LaskyLun07,Gautreau:1984PhRvD186}:
\begin{gather}
\ud s^2 = - \alpha(t,r)^2 \ud t^2 + \frac{(-\beta(t,r) \ud t + \ud r)^2}{1 + E(t,r)} + r^2 \ud \Omega^2 \,. \label{eq:GPGmetric}
\end{gather}


In order to interpret the meaning of $\beta(t,r)$, we compute $n^a$ in GPG coordinates
\begin{gather}
n^a \rightarrow \frac{1}{\alpha(t,r)} \left( \partial_t + \beta(t,r) \partial_r \right)\,,
\end{gather}
which is given by
\begin{gather}
n^a \partial_a r = \frac{\beta(t,r)}{\alpha(t,r)}\,. \label{eq:partialnr}
\end{gather}

Combining Eqs.~\eqref{eq:M_ms_develop} and \eqref{eq:partialnr}, we obtain for $\beta$:
\begin{gather}
    \frac{\beta(t,r)}{\alpha(t,r)}=\pm\sqrt{\frac{2M_{ms}(t,r)}{r}+E(t,r)}\,,\label{eq:betaDef}
\end{gather}
\noindent
which we interpret as the radial velocity of the flow, where the $'-'$ (minus) sign corresponds to collapse and the $'+'$ (plus) sign to expansion\footnote{Note that our definition for $\beta(t,r)$ adopts the opposite sign to that of  
Ref.~\cite{Lasky:2006zz} 
}.

\section{Physical interpretations}\label{sec:PhysInterp}
\subsection{Interpreting the flow equations}\label{Sec:Interpret}

The Raychaudhuri equations \eqref{eqs:RayNE} encode the evolution and constraint equations for spacetime and matter, given the symmetry assumptions. In order to render 
explicit these relations we will show that Eq.~\eqref{eq:RayNE-a} is equivalent to the generalised TOV (gTOV) functional \cite{Mimoso:2013iga} that provides an effective acceleration of 
matter shells in this spacetime. We assume in this section that matter is 
adequately modeled 
by a perfect fluid, with energy-momentum tensor given by
\begin{gather}
T_{ab} = \rho n_a n_b + P (e_a e_b + s_{ab})\,,
\end{gather}
where $\rho$ and $P$, respectively, are the energy-density and the isotropic pressure measured by the observer comoving with velocity $n^a$.
Thus we have 
\begin{gather}
T_{ab}n^a n^b = \rho\,, \quad T_{ab}e^a e^b = P\,, \quad T_{ab}n^a e^b = 0\,.\label{eq:perfectfluid}
\end{gather}
and, by applying the conservation of the energy-momentum $\nabla_a T^{ab} = 0$ along the $n^a$ and $e^a$ directions we derive the equations 
\begin{subequations}
\begin{align}
 n^a \nabla_a \rho &+ (\rho + P) (\mathcal{B} + \Theta_n) = 0\,, \label{eq:continuity}\\
\mathcal{A} &= -\frac{e^a \nabla_a P}{\rho + P}\,, \label{eq:accel}
 \end{align}  
\end{subequations}
\noindent
which relate the $1+1+2$ scalars to the latter hydrodynamic variables. 
These Eqs.~(\ref{eq:continuity}) and (\ref{eq:accel}) complement 
Eqs.~\eqref{eqs:RayNE} to form the set of field equations. 

Before proceeding it is appropriate to write up the anticipated relation between the scalar quantities $\mu$ and $Q$, associated with the dual null directions $k^a$ and $l^a$, and the thermodynamic quantities (\ref{eq:perfectfluid}). Indeed, 
notice that $\mu$ and $Q$ in 
Eqs.~\eqref{eqs:Raynull} reduce to $\mu=T_{cd}k^c k^d=\frac{\rho+P}{4}$, which appears related to the fluid enthalpy $h=E/V+P$,  and $Q=T_{ab}k^al^b=\frac{\rho-P}{4}$, that could be interpreted as related to a fluid enthalpy for negative pressure $"h"=E/V+(-P)$, so that, as expected the null energy condition $\mu\ge 0$ translates into $\rho+P\ge 0$.\footnote{Incidentally, if $T_{ab}$ coincides with the EMT of a cosmological constant the null energy condition is marginally satisfied, since $\rho+P=0$.}


We remark that the Misner-Sharp mass can be related to the 2-expansions as
\begin{gather}
- \Theta_{(n)}^2 + \Theta_{(e)}^2 = \frac{4}{r^2} \left(1 - \frac{2M_{ms}}{r}\right)\,, \label{eq:ThetaMS}
\end{gather}
\noindent
which, applied to Eq.~\eqref{eq:RayNE-a} along with~\eqref{eq:perfectfluid} leads to
\begin{gather}
\Lie_n \Theta_{(n)} = - \frac{\Theta_{n}^2}{2} + \frac{1}{r^2} \left(1 - \frac{2M_{ms}}{r}\right) +\frac{1}{r^2} +\mathcal{A}\Theta_{(e)} - 8\pi P\,,
\end{gather}
\noindent
which can be further simplified by using Eq.~\eqref{eq:accel} and 
\begin{gather}
e^a \partial_a = \sqrt{1+E}\partial_r\,,
\end{gather}
\noindent leading to
\begin{gather}
\Lie_n \Theta_{(n)} + \frac{\Theta_{(n)}^2}{2} = - \frac{2}{r}\left( \frac{1+E}{\rho + P} \partial_r P + \frac{M_{ms}}{r^2} + 4\pi P r\right) = -\frac{2}{r} \text{gTOV}\,,
\end{gather}
\noindent which was already shown in \cite{Maciel:2015vva}, although here the result is obtained in a more straightforward procedure.

We can also see that the Raychaudhuri equations along $e^a$, Eqs~\eqref{eq:RayNE-b} and~\eqref{eq:RayNE-d} are constraints equations. By computing 
\begin{gather}
\Lie_e (- \Theta_{(n)}^2 + \Theta_{(e)}^2 ) = - 2\Theta_{(n)} \Lie_e \Theta_{(n)} + 2 \Theta_{(e)}\Lie_e \Theta_{(e)}\,,
\end{gather}
we find, after using Eqs.~\eqref{eq:RayNE-b}, \eqref{eq:RayNE-d}, \eqref{eq:perfectfluid} and \eqref{eq:ThetaMS} and some manipulation:
\begin{gather}
\Lie_e M_{ms} = 2 \Theta_{(e)} \rho r^3\,,
\end{gather}
which, using the translation to GPG coordinates, Eq.~\eqref{eq:etranslationup}, leads to
\begin{gather}
\partial_r M_{ms} = 4\pi \rho r^2\,,
\end{gather}
\noindent
as expected from the mass-energy definition.



\subsection{Flow scalars from 2D to 3D}\label{sec:flow2}

In order to compute the extrinsic curvature of the spatial sections orthogonal to the flow $n^a$, we use the definition
\begin{gather}
K_{ab} = \frac{1}{2}\Lie_n h_{ab} = \frac{1}{2}\Lie_n (e_a e_b + s_{ab})\,.
\end{gather}
The easiest way to translate this equation in terms of the dual null flow scalars is to project it
by making use of the symmetries:
\begin{gather}
    s^{ab}K_{ab} =  \Theta_n\,, \quad n^a K_{ab} = 0\,, \quad e^a K_{ab} = \frac{1}{2}\left( \Lie_n e_b + e^a e_b \Lie_n e_a \right) = \mathcal{B} e_b\,,
\end{gather}
which results in 
\begin{gather}
K_{ab} = \mathcal{B}e_a e_b + \frac{\Theta_n}{2} s_{ab}\,.
\end{gather}

From the extrinsic curvature, we compute the 3D expansion $\Theta$:
\begin{gather}
\Theta = g^{ab}\nabla_a n_b = K\indices{_a ^a} = \mathcal{B} + \Theta_n\, .\label{eq:theta3}
\end{gather}
Since  the shear tensor $\sigma_{ab}$ is defined as the trace free part of $K_{ab} = h\indices{_a^c}h\indices{_b^d}\nabla_c n_d$, we obtain 
\begin{gather}
\sigma_{ab} = \mathcal{B}e_ae_b + \frac{\Theta_n}{2} s_{ab} - \frac{1}{3}\left(\mathcal{B} +\Theta_n\right) \left(e_a e_b + s_{ab}\right) = \nonumber\\
= \frac{\Theta_n - 2\mathcal{B} }{6}\left(-2e_a e_b + s_{ab}\right)\,, \label{ssecB_shear} 
\end{gather}
\noindent
where we recognize the symmetric trace free projector (STFPr) $P_{ab}= -2e_a e_b + s_{ab}$, (check Ref.\cite{Maciel:2015vva}). In the latter equation (\ref{ssecB_shear}) we define the shear scalar as \begin{gather}
  \sigma = \frac{\Theta_n - 2\mathcal{B} }{6} \,,
\end{gather}
which, combined with Eq.~\eqref{eq:theta3}, leads to the relation between 2D and 3D expansions and shear
\begin{subequations} \label{eq:ThetaandB}
\begin{align}
      \Theta_n &= \frac{2\Theta}{3} + 2\sigma \ , \label{eq:Thetameaning}\\
\mathcal{B} &= \frac{\Theta}{3} - 2\sigma\,, \  \label{eq:Bmeaning}
\end{align}
\end{subequations}
\noindent
that shows that the pair of scalars $\Theta_n$ and $\mathcal{B} = K_{ab}e^ae^b$ is equivalent to the pair $\Theta$ and $\sigma$ in order to describe the flow in spherical symmetry.

\subsection{Redshift}\label{sec:redshift}

To compute the redshift suffered by a photon travelling radially, we have to write the wave vector $K^a$ in our formalism. Since the frequency $\omega$ measured by the observer whose four velocity is $n_a$ is defined as
\begin{gather}
\omega = - K^a n_a\,,
\end{gather}
\noindent
and the wave vector is null, 
it can be written as
\begin{gather}
K^a = \omega \left( n^a \pm e^a \right)\,,\label{eq:wavevector}
\end{gather}
\noindent
provided that the wave vector satisfies the geodesic equation, $K^a \nabla_a K_b = 0$. The geodesic equation applied to \eqref{eq:wavevector} gives us
\begin{gather}
    \omega (n^a \pm e^a) \nabla_a \left[\omega(n_b \pm e_b)\right] = 0\,,
\end{gather}
\noindent
which, after using the definitions \eqref{eq:AandB}, implies the following equality
\begin{gather}
(n^a \pm e^a) \frac{\nabla_a \omega}{\omega} = \mp \mathcal{A} - \mathcal{B} \,,
\end{gather}
\noindent
that we use to write 
\begin{gather}
\nabla_a \ln \omega = \mathcal{B}n_a - \mathcal{A}e_a\,. \label{eq:delomega}
\end{gather}
By defining the redshift as $1+z = \frac{\omega}{\omega_0}$, where $\omega_0$ is the emitted frequency, we obtain the relation
\begin{gather}
\Lie_n \ln{(1+z)} = - \mathcal{B}\,. \label{eq:redshift}
\end{gather}

Eq.~\eqref{eq:redshift} gives us the interpretation of the $\mathcal{B}$ scalar as the source for redshift for general spherically symmetric spacetimes. Note that by Eq.~\eqref{eq:Bmeaning}, $\mathcal{B}$ corresponds to a linear combination of the 3-dimensional expansion $\Theta$ and 3-dimensional shear scalar $\sigma$. It geometrically encapsulates previous expressions \cite{Ellis:1971pg,Plebanski:2006sd}, and for an isotropic and homogeneous model it just corresponds to the Hubble constant $H$, hence returning the well know relation $1+z = \frac{a_0}{a}$. 

Using GPG coordinates, Eq.~\eqref{eq:redshift} acquires the form
\begin{align}
\frac{1}{\alpha}\left(\partial_t + \beta \partial_r \right) \ln(1+z) =&  -\frac{1}{\alpha} \frac{\beta \partial_r E +\partial_t E}{2(1+E)} + \frac{\partial_r \beta}{\alpha}\nonumber\\
=& -\frac{\mathcal{L}_{n}E}{2\left(1+E\right)}-\frac{\beta^{\prime}}{\alpha}\,,
\end{align}
\noindent in which we recognise the left hand side of the momentum constraint from the EFE. 

That momentum constraint in GPG coordinates reads 
\begin{align}
-\frac{\mathcal{L}_{n}E}{2\left(1+E\right)}-\frac{\beta^{\prime}}{\alpha} & =\left(\mathcal{L}_{n}r\right)^{\prime},\label{eq:MomentumC}
\end{align}
 and we recall the GPG expressions for the 3-expansion and 3-shear
scalars 
\begin{align}
\Theta= & \left(\mathcal{L}_{n}r\right)^{\prime}+2\frac{\mathcal{L}_{n}r}{r},\label{eq:ThetaInRj}\\
\sigma= & -\frac{1}{3}\left[\left(\mathcal{L}_{n}r\right)^{\prime}-\frac{\mathcal{L}_{n}r}{r}\right].\label{eq:ShearInRj}
\end{align}
 Therefore, the redshift source can be recast to relate it evidently
to the 3-expansion and 3-shear as 
\begin{eqnarray}
\frac{1}{\alpha}\left(\partial_{t}+\beta\partial_{r}\right)\ln(1+z) & = & \left(\mathcal{L}_{n}r\right)^{\prime}\;,\nonumber \\
 & = & \frac{1}{3}\left[\left(\mathcal{L}_{n}r\right)^{\prime}+2\frac{\mathcal{L}_{n}r}{r}\right]+\frac{2}{3}\left[\left(\mathcal{L}_{n}r\right)^{\prime}-\frac{\mathcal{L}_{n}r}{r}\right]\;,\nonumber \\
 & = & \frac{\Theta}{3}-2\sigma\;.
\end{eqnarray}


In 
the restricted case of the Lemaître-Tolman-Bondi (LTB) dust  model,  we get
\[
 \partial_t \ln(1+z) = - \partial_t \ln R'
\]
so that we are led to 
\[
1+z = \frac{R'_0}{R'} \;
\]
(See for instance ~\cite{Codur:2021wrt}.) 

Notice that this fits again our expression for $\mathcal{B}$, as
\[
\frac{\dot R'}{R'} = \frac{1}{3}\left(2\frac{\dot R}{R}+\frac{\dot R'}{R'}\right) -2\,\frac{1}{3}\,\left( \frac{\dot R}{R}-\frac{\dot R'}{R'}\right) =\Theta/3-2\sigma.
\]


\section{Extensions and comparisons}\label{sec:extensionsComp}
\subsection{Hyperbolic and planar symmetries} \label{sec:hyperbolic}

The 1+1+2 treatment is also convenient for spacetimes with codimension-two hyperbolic and rigid planar foliations. We can express spacetimes of the three types of symmetries, spherical, planar and hyperbolic in one unified notation using
\begin{gather}
s_{ab}\ud x^a \ud x^b = r^2 \left( \ud \theta^2 + S^{2}_\epsilon (\theta) \ud \phi^2 \right) \equiv r^2 \ud \Sigma^2 \,,
\end{gather}
\noindent
where
\begin{gather}
S_\epsilon(\theta) = \left\{ \begin{tabular}{c l}
 $\sin \theta\,,$    &  for $\epsilon = 1\,,$  \\
  $\theta\,,$   &      for $\epsilon = 0\,,$  \\
  $\sinh \theta\,,$ & for $\epsilon = -1\,.$ 
\end{tabular}
\right.
\end{gather}

It is worth noting that the quantities $r$, $\theta$ and $\phi$ do not keep their intuitive geometrical meanings in the cases of planar and hyperbolic symmetry. Another important remark is that the planar symmetry considered here is not the general case,
but only that associated with the "rigid" planar symmetric spacetimes, which we can define as the cases where the 2-shear, defined as, $\sigma^{(v)}_{ab}= s\indices{_a ^c}s\indices{_b ^d} \nabla_c v_d - \frac{s_{ab}}{2} \Theta_{(v)}$ vanishes for any vector $v^a$ generated by $( n^a \,, e^a )$.

We can generalize Eqs.~\eqref{eqs:RayNE} for the three symmetries as
\begin{subequations}
    \begin{align}
       \Lie_n \Theta_{(n)} &= - \frac{3\Theta_{(n)}^2}{4} + \frac{ \Theta_{(e)}^2}{4} + \frac{\epsilon}{r^2}+ \mathcal{A}\Theta_{(e)} - 8 \pi T_{ab}e^a e^b \,, \\
 \Lie_e \Theta_{(e)} &= -\frac{3\Theta_{(e)}^2}{4} + \frac{\Theta_{(n)}^2}{4} + \mathcal{B} \Theta_{(n)} - \frac{\epsilon}{r^2} - 8\pi T_{ab}n^a n^b\,,\\
 \Lie_n \Theta_{(e)} &=  - \frac{\Theta_{(n)} \Theta_{(e)}}{2} + \mathcal{A} \Theta_{(n)}  - 8 \pi T_{ab}n^a e^b \,, \\
 \Lie_e \Theta_{(n)} &= - \frac{\Theta_{(n)}\Theta_{(e)}}{2} + \mathcal{B} \Theta_{(e)} - 8 \pi T_{ab}n^a e^b\,,
    \end{align}
\end{subequations}
\noindent
by following the same procedure we used for treating the spherically symmetric case.

We can also generalize Eq.~\eqref{M_ms-def1} by using the generalization of Hawking-Hayward mass-energy (and by extension, of the Misner-Sharp mass-energy) proposed in \cite{Maciel_2020} in order to obtain
\begin{gather}
g^{ab}\partial_a r \partial_b r = \epsilon - \frac{2\mu}{r}\,, \label{eq:massparameter}
\end{gather}
\noindent
where $\mu$ corresponds to the mass parameter in the planar and hyperbolic cases, similar to a superficial mass density, as explained in \cite{Maciel_2020} and it corresponds the Misner-Sharp mass-energy in the spherical case. Using Eq.~\eqref{eq:massparameter} and following the procedure of Secs.~\ref{sec:GPG} and \ref{Sec:Interpret}, we obtain

\begin{gather}
\ud s^2 = - \alpha^2 \ud t^2 + \frac{(-\beta \ud t + \ud r)^2}{\epsilon + E} + r^2 \ud \Sigma^2\,,\\
\text{gTOV} = \frac{\epsilon+E}{\rho + P} \partial_r P + \frac{M_{ms}}{r^2} + 4\pi Pr\,.\label{eq:generalgTOV}\\
\partial_r \mu = 4\pi r^2 \rho\,.
\end{gather}

When the solution is static, $\Theta_{(n)}=0$, which corresponds to $E = \frac{2\mu}{r}$ in GPG language, Eq.~\eqref{eq:generalgTOV} reduces to the generalized TOV equation analysed in Ref.~\cite{Maciel_2020}.

The results of Secs.~\ref{sec:flow2} and~\ref{sec:redshift} depend only on the warped symmetric foliation and are valid for planar and hyperbolic symmetries without changes, including Eqs.~\eqref{eq:ThetaandB} and~\eqref{eq:redshift}.

\subsection{Advantages of 1+1+2/2+2 over Newman-Penrose formalism} \label{sec:advantages}
The Raychaudhuri equations correspond to the trace of the evolution of the $k$ or $l$ congruences covariant derivatives, while these covariant derivatives are themselves represented as a series of transport equations in the NP formalism \cite{Sachs:1964zza,Newman:1961qr}. The complexity of obtaining the Raychaudhuri equations in NP formalism appears therefore higher than using the optical scalars in the 2+2 approach, demonstrating its better fit to our domain of interest. 

NP formalism uses the entirely null complex tetrad to represent the metric as
\begin{subequations}
    \begin{align}
z_{A}^{a}= & \left(l_{NP}^{a},n_{NP}^{a},m^{a},\overline{m}^{a}\right),\\
\eta_{AB}=\eta^{AB}= & \left(\begin{array}{cccc}
0 & -1 & 0 & 0\\
-1 & 0 & 0 & 0\\
0 & 0 & 0 & 1\\
0 & 0 & 1 & 0
\end{array}\right),\\
\Rightarrow g^{ab}= & \eta^{AB}z_{A}^{a}z_{B}^{b},
\end{align}
\end{subequations}
the NP spin coefficients, taking $l_{NP}=\sqrt{2}k$ and $n_{NP}=\sqrt{2}l$ to correct normalisation, that account for projected parallel transports of the tetrad members, involved in transport equations, that regulate their parallel transports and scalars for projections of the Weyl and Ricci tensors. 

We 
use the following definitions for spin coefficients,
\begin{subequations}  
\begin{align}
\varrho\equiv & -\sqrt{2}m^{a}\overline{m}^{b}\nabla_{b}k_{a}, & \varkappa\equiv & -2m^{a}k^{b}\nabla_{b}k_{a}, & \varepsilon\equiv & \frac{1}{\sqrt{2}}\left(\overline{m}^{a}k^{b}\nabla_{b}m_{a}-2l^{a}k^{b}\nabla_{b}k_a\right),\\
\mu\equiv & \sqrt{2}\overline{m}^{a}m^{b}\nabla_{b}l_{a}, & \nu\equiv & 2\overline{m}^{a}l^{b}\nabla_{b}l_{a}, & \gamma\equiv & \frac{1}{\sqrt{2}}\left(\overline{m}^{a}l^{b}\nabla_{b}m_{a}-2l^{a}l^{b}\nabla_{b}k_{a}\right),\\
 &  & \pi\equiv & 2\overline{m}^{a}k^{b}\nabla_{b}l_{a}, & \tau\equiv & -\sqrt{2}m^{a}n^{b}\nabla_{b}k_{a},
\end{align}
\end{subequations}
for transport equations,
\begin{subequations}
    \begin{align}
D\equiv & \sqrt{2}k^{a}\nabla_{a}, & \Delta\equiv & \sqrt{2}l^{a}\nabla_{a},\\
D\sqrt{2}k= & \left(\varepsilon+\overline{\varepsilon}\right)\sqrt{2}k-\overline{\varkappa}m-\varkappa\overline{m}, & D\sqrt{2}l= & -\left(\varepsilon+\overline{\varepsilon}\right)\sqrt{2}l+\pi m+\overline{\pi m},\\
\Delta\sqrt{2}k= & \left(\gamma+\overline{\gamma}\right)\sqrt{2}k-\overline{\tau}m-\tau\overline{m}, & \Delta\sqrt{2}l= & -\left(\gamma+\overline{\gamma}\right)\sqrt{2}l+\nu m+\overline{\nu m},
\end{align}
\end{subequations}
and for Ricci scalars
\begin{subequations}
    \begin{align}
\Phi_{00}\equiv & R_{ab}k^{a}k^{b}\vphantom{\frac{R_{ab}1}{2R_{ab}}}, & \Phi_{22}\equiv & R_{ab}l^{a}l^{b}\vphantom{\frac{R_{ab}1}{2R_{ab}}}, & \Phi_{11}\equiv & \frac{1}{4}R_{ab}\left(2k^{a}l^{b}+m_{a}\overline{m}_{b}\right)\vphantom{\frac{R_{ab}1}{2R_{ab}}},\\
 & \textrm{ and} & \Phi_{11}+3\varLambda= & R_{ab}k^{a}l^{b},\textrm{ as} & \varLambda\equiv & \frac{R}{4!}\vphantom{\frac{R_{ab}1}{2R_{ab}}}=-2\left(2k_{a}l_{b}-m_{a}\overline{m}_{b}\right)\frac{R^{ab}}{4!}.
\end{align}
\end{subequations}
We then can link the NP spin coefficients to the 2+2 inaffinities as
\begin{subequations}
   \begin{align}
k.\Delta k= & 0, & l.Dk= & \left(\varepsilon+\overline{\varepsilon}\right)k.l\equiv\sqrt{2}\nu_{k}k.l,\\
l.Dl= & 0, & k.\Delta l= & -\left(\gamma+\overline{\gamma}\right)k.l\equiv\sqrt{2}\nu_{l}k.l.
\end{align}
\end{subequations}
We can define the expansions in NP formalism as
\begin{align}
\Theta_{\left(k\right)}= & \nabla_{a}k^{a}=g^{ab}\nabla_{a}k_{b}\nonumber \\
= & \eta^{AB}z_{A}^{a}z_{B}^{b}\nabla_{a}k_{b}\nonumber \\
= & \left[-2k^{a}l^{b}-2l^{a}k^{b}+\left(m^{a}\overline{m}^{b}+\overline{m}^{a}m^{b}\right)\right]\nabla_{a}k_{b}\nonumber \\
= & -2\left(k^{a}l^{b}+l^{a}k^{b}\right)\nabla_{a}k_{b}+\left(m^{a}\overline{m}^{b}+\overline{m}^{a}m^{b}\right)\nabla_{a}k_{b}\nonumber \\
= & -\sqrt{2}\left(l^{b}Dk_{b}+k^{b}\Delta k_{b}\right)-\frac{\left(\varrho+\overline{\varrho}\right)}{\sqrt{2}}=\frac{\left(\varepsilon+\overline{\varepsilon}\right)-\left(\varrho+\overline{\varrho}\right)}{\sqrt{2}},\\
\Theta_{\left(l\right)}= & -\sqrt{2}\left(l^{b}Dl_{b}+k^{b}\Delta l_{b}\right)+\left(m^{a}\overline{m}^{b}+\overline{m}^{a}m^{b}\right)\nabla_{a}l_{b}=\frac{\mu+\overline{\mu}-\left(\gamma+\overline{\gamma}\right)}{\sqrt{2}}.
\end{align}

The evolution traces left-hand-sides can then be written as
\begin{subequations}
    \begin{align}
k^{a}\nabla_{a}\Theta_{\left(k\right)}= & \frac{D\Theta_{\left(k\right)}}{\sqrt{2}}, & k^{a}\nabla_{a}\Theta_{\left(l\right)}= & \frac{D\Theta_{\left(l\right)}}{\sqrt{2}},\\
l^{a}\nabla_{a}\Theta_{\left(k\right)}= & \frac{\Delta\Theta_{\left(k\right)}}{\sqrt{2}}, & l^{a}\nabla_{a}\Theta_{\left(l\right)}= & \frac{\Delta\Theta_{\left(l\right)}}{\sqrt{2}}.
\end{align}
\end{subequations}
Using the 2+2 approach in Eqs.~\eqref{eqs:Raynull},  and the sources in the language of NP formalism, the Raychaudhuri equations in  spherical symmetry can be expressed as
\begin{subequations}
    \begin{align}
D\left(\frac{\left(\varepsilon+\overline{\varepsilon}\right)-\left(\varrho+\overline{\varrho}\right)}{2}\right)= & \frac{\left(\left(\varepsilon+\overline{\varepsilon}\right)-\left(\varrho+\overline{\varrho}\right)\right)^{2}}{4}-\Phi_{00},\\
\Delta\left(\frac{\mu+\overline{\mu}-\left(\gamma+\overline{\gamma}\right)}{2}\right)= & \frac{\left(\mu+\overline{\mu}-\left(\gamma+\overline{\gamma}\right)\right)^{2}}{4}-\Phi_{22},\\
\Delta\left(\frac{\left(\varepsilon+\overline{\varepsilon}\right)-\left(\varrho+\overline{\varrho}\right)}{2}\right)= & -\frac{\left(\left(\varepsilon+\overline{\varepsilon}\right)-\left(\varrho+\overline{\varrho}\right)\right)\left(\mu+\overline{\mu}-2\left(\gamma+\overline{\gamma}\right)\right)}{2}-\frac{k^{a}l_{a}}{r^{2}}+\Phi_{11}-3\varLambda,\\
D\left(\frac{\mu+\overline{\mu}-\left(\gamma+\overline{\gamma}\right)}{2}\right)= & -\frac{\left(2\left(\varepsilon+\overline{\varepsilon}\right)-\left(\varrho+\overline{\varrho}\right)\right)\left(\mu+\overline{\mu}-\left(\gamma+\overline{\gamma}\right)\right)}{2}-\frac{k^{a}l_{a}}{r^{2}}+\Phi_{11}-3\varLambda.
\end{align}
\end{subequations}
The NP formalism does not provide an easy way to compute the right-hand-side of the evolution traces, showing the 2+2 approach to be better fit to tackle the problems under consideration, namely those concerning the interplay between matter and lightcones, as well as observer centered radiation problems.

\section{Conclusion}
This work started by reviewing the fundamentals of the dual null formalism, its projector on 2-space, the optical geometric decomposition of its kinematic and its dynamics, reduced in spherical symmetry to the Raychaudhuri equations. 

The complete derivation including the cross focusing equation is presented for the fir
st time, as far as we know, in the  appendices.

We proposed a dictionary to translate between this splitting and the 1+1+2 approach in a structured way, in order to highlight the physical interpretation of the flow scalars that arise from this formalism, as well as connect it to the coordinated dependent GPG formalism and the 1+3 splitting of the spacetime. 

Such dictionary clearly relates Raychadhuri-like equations to the dynamics of the flow or null fields and present transparently the role of flow acceleration and the shape tensor associated with the foliation on the dynamics. Such clear spelling can prove useful for the fields interested in the interplay between lightcones and matter flow, as well as observer centered point of view to radiation problems, such as Gravitational wave effects on environment. 

Physically, we recalled that the flow equation relates to a generalised TOV hydrostatic equilibrium, while the cross focusing and radial Raychaudhuri equations produce the Misner-Sharpe mass conservation. We transparently spelled out the interpretation of 1+1+2 scalars as combination of flow 3-expansion and -shear, as well as an interesting connection to redshift that unveils that the $\mathcal{B}$ scalar, which appears naturally on the 1+1+2 evolution equations, is also the source of the redshift in warped symmetric spacetimes. To the best of our knowledge, the gauge invariant scalar $\mathcal{B}$ is introduced for the first time. This is a strong argument in favor of the realisation that the 1+1+2 formalism deals with the physically meaningful scalars in the context of matter dominated warped symmetric spacetimes. 

Finally we addressed the generalisation of the above framework from the symmetry of spherical foliations to planar and hyperbolic foliations of the spacetime. We presented a thorough comparison showing the advantage of using the present framework rather than the more convoluted 
NP formalism in such problems. This work should render obvious the interest of the presented framework for future explorations of  warped, foliated spacetimes .

\section*{Acknowledgements}
JPM acknowledges funding from the Funda\c c\~{a}o para a Ci\^encia e a Tecnologia (FCT, Portugal) under research projects UIDB/04434/2020, UIDP/04434/2020, EXPL/FIS-AST/1368/2021 and PTDC/FIS-AST/0054/2021. MLeD acknowledges the financial support by the Lanzhou University starting
fund and the Fundamental Research Funds for the Central Universities
(Grant No. lzujbky-2019-25).

\appendix

\section{Raychaudhuri equation for null, foliatedly symmetric, congruences}\label{AppRay}

Our aim in this section is to compute $\Lie_{k} \Theta_{(k)}$. It is important to define the quantities $\nu_k$ and $\nu_l$ as
\begin{gather}
\nabla_a k_b = \frac{1}{k^c l_c}\left(\nu_k {l_a k_b} - \nu_l {k_a k_b}\right) + \frac{\Theta_{(k)}}{2} s_{ab}\,, \label{nablak}
\end{gather}
\noindent where the last term comes from the definition of expansion, Eq.~\eqref{expansion}. Since we keep $k^a l_a$ constant, we have $l^a \nabla k_a = - k^a \nabla l_a$, therefore
\begin{gather}
\nabla_a l_b =\frac{1}{k^c l_c}\left( \nu_l {k_a l_b} - \nu_k l_a l_b \right) + \frac{\Theta_{(l)}}{2} s_{ab}\,. \label{nablal}
\end{gather}

Those quantities are interpreted as the \textit{inaffinities} of the vectors $k^a$ and $l^a$, since, 
\begin{gather}
k^a \nabla_a k_b = \nu_k k_b\,, \quad \text{and} \quad l^a \nabla_a l_b = \nu_l l_b\,,
\end{gather}
\noindent which means that when $\nu_k = \nu_l = 0$ the dual null basis $(k^a, l^a)$ corresponds to a geodesic null field.
Consider the $k$-directed covariant derivative of $ \nabla_{a}k_{b}$:
\begin{gather}
  k^{c}\nabla_{c} \nabla_{a}k_{b} =
 R\indices{_c _a _b ^d}k^{c}k_{d} +k^{c} \nabla_{a} \nabla_{c} k_{b}=
  \nonumber\\
 = - R_{c a d b} k^{c} k^{d} + k^{c} \nabla_{a} \left[ \frac{1}{k^c l_c}( \nu_k l_c k_b - \nu_l k_c k_b) + \frac{\Theta_{(k)}}{2}s_{cb}\right] = \nonumber\\
 = - R_{c a d b} k^{c} k^{d} + \nu_k \nabla_{a}k_b + \frac{\Theta_{(k)}}{2}k^c \nabla_a s_{cb} + (\dots)_a k_b \,,
 \label{KDDK}
\end{gather}
\noindent
where $R\indices{_a_b_c^d}$ is the Riemann curvature tensor and we gathered under $(\dots)_a k_b$ all terms that are $k$ 
directed in the second index. We have also used the symmetry of the Riemann tensor and the fact that  $k^c s_{cb} = 0$.

Thus, contracting \eqref{KDDK} with $s^{a b}$, we obtain
\begin{gather}
s^{ab} k^{c}\nabla_{c} \nabla_{a}k_{b} = - s^{ab}R\indices{ _ c _a _d _b}k^c k^d + \nu_k s^{ab}\nabla_a k_b + \frac{\Theta_{(k)}}{2}s^{ab}k^c \nabla_a s_{bc} \Rightarrow\nonumber\\
s^{ab} k^{c}\nabla_{c} \nabla_{a}k_{b} = -R\indices{ _ c _m _d ^m}k^{c}k^{d} + \nu_k \Theta_{(k)} -  \frac{\Theta_{(k)}}{2} s^{ab}s\indices{ ^{c} _{b}} \nabla_a k_c \Rightarrow \nonumber\\
 s^{ab} k^{c}\nabla_{c} \nabla_{a}k_{b}= -R_{cd}k^{c}k^{d}  + \nu_k \Theta_{(k)}- \frac{\Theta_{(k)}^2}{2} \,, \label{Raychaudhuri-p1}
\end{gather}
\noindent
where $R_{ab}$ is the Ricci tensor and we used $s^{ab}R_{cadb}k^c k^d = g^{ab}R_{cabd}k^c k^d$, since $R\indices{_{c} _{a} _{b} _{d}}k^{(a}l^{b)}k^c k^d = 0$ due to the symmetries of the Riemann tensor. 

On the other hand, we can express the left hand side of Eq.~\eqref{Raychaudhuri-p1} as
\begin{gather}
s^{ab} k^{c}\nabla_{c} \nabla_{a}k_{b} = k^c \nabla_c \left(s^{ab}\nabla_a k_b \right) - \nabla_a k_b k^c \nabla_c s^{ab} \Rightarrow \nonumber\\
s^{ab} k^{c}\nabla_{c} \nabla_{a}k_{b} = \Lie_k \Theta_{(k)} - \nabla_a k_b k^c \nabla_c s^{ab} = \Lie_k \Theta_{(k)}\,, \label{eq:LiekThetak}
\end{gather}
since, by Eq.~\eqref{nablak}, we have
\begin{gather}
\nabla_a k_b k^c \nabla_c s^{ab} = \left[\frac{1}{k^c l_c}\left(\nu_k {l_a k_b} - \nu_l {k_a k_b}\right) + \frac{\Theta_{(k)}}{2} s_{ab}\right] k^c \nabla_c s^{ab} \Rightarrow \nonumber\\
\nabla_a k_b k^c \nabla_c s^{ab} = -\frac{1}{k^c l_c}\left(\nu_k {l_a s^{ab}} - \nu_l {k_a s^{ab}}\right)k^c \nabla_c k_b + \frac{\Theta_{(k)}}{2} s_{ab}k^c\nabla_c s^{ab} = 0\,, \label{nablakSab}
\end{gather}
where we used the fact that $k_b s^{ab}$ and $s^{ab}s_{ab}$ are both constants in order to manipulate the derivatives in the Eq.~\eqref{nablakSab}.
Generalizing for both null directions and using the Einstein's equations:
\begin{gather}
 R_{ab}-\frac{1}{2}R g_{ab} = 8 \pi  T_{ab}\,.
\end{gather}
\noindent we finally obtain
\begin{gather}
 \Lie_{k}\Theta_{(k)} =  - \frac{\Theta_{(k)}^2}{2} +  \nu_k \Theta_{(k)} - 8\pi T_{cd}k^c k^d. \label{Raychaudhuri} 
\end{gather}
\noindent

\section{The cross focusing equation}\label{AppCross}

In order to deduce the other evolution equation, relating the change of one expansion as we move along the \emph{other} null direction, namely $\Lie_{k} \Theta_{(l)}$ and $\Lie_l \Theta_{(k)}$. They are closely related since
\begin{gather}
    \Lie_{k} \Theta_{(l)} = \Lie_{k} \left( \frac{2}{r}\Lie_{l} r \right) = \Lie_{k} \Lie_{l} \ln r^2 = 
   \left(\Lie_l \Lie_k + \Lie_{[k,l]} \right) \ln r^2 =\nonumber\\ 
   = \Lie_l \Theta_{(k)} + \Theta_{([k,l])} = \Lie_l \Theta_{(k)} - \nu_k \Theta_{(l)} + \nu_l \Theta_{(k)}\,,
\end{gather}
\noindent where $[k,l]^c = \Lie_k l^c = \nu_l k^c - \nu_k l^c$ is the commutator between the vectors $k^a$ and $l^a$.

It will be convenient at this point to evoke the existence of coordinate null basis along our $k^a$ and $l^a$ directions, which means a pair of commuting vectors parallel to this direction. We denote those vectors as $\delta^{a}_\alpha$, where we the Greek indices denote coordinate labels instead of tensor indices. We also define the covariant vectors $\delta_a^{\alpha}$ by the relation $\delta^a_{\alpha}\delta_a^{\beta} = \delta_{\alpha}^{\beta}$, where in the right hand side we interpret $\delta$ as the Krönecker symbol. In particular, for the null directions $\delta^a_+$ and $\delta^a_-$, respectively, defined such that
\begin{gather}
[\delta^a_+, \delta^b_-] = 0\,, \quad k^a = K \delta^a_+ \, \quad l^a= L \delta^a_-\,, 
\end{gather}
\noindent
where $K$ and $L$ are scalar functions that can vary only along $k^a$ and $l^a$ directions, that is $s^{ab}\nabla_b K = s^{ab}\nabla_b L =0$, due to spherical symmetry. The functions $K$ and $L$ which relate the original dual null basis with a coordinate basis must satisfy:
\begin{gather}
[ \delta^a_+ , \delta^b_-] = [K^{-1} k^a, L^{-1}l^b] = 0 \,, \quad \text{which implies} \nonumber\\
K \Lie_k \ln L = - \nu_k \,, \quad \text{and} \quad L \Lie_l \ln K = -\nu_l\,.
\end{gather}

We define the metric components $g_{\alpha \beta} = g_{ab}\delta^a_{\alpha} \delta^b_{\beta}$, in particular, $g_{+-} = g_{ab}\delta^a_+ \delta^b_- = -e^f$, where $f$ is also a function that only varies along directions orthogonal to the spheres of symmetry, so the metric tensor may be written as:
\begin{gather}
 g_{ab} = -2 e^f \delta^+_{(a}\delta^-_{b)} + s_{\alpha \beta} \delta^{\alpha}_{a} \delta^{\beta}_{b}\,. \label{eq:metric-coordinates}
\end{gather}
Therefore, we obtain for the covariant vectors:
\begin{gather}
 k_{a} = g_{ab}k^b = -Ke^f\delta_a^{-}\,, \quad l_{a} = g_{ab}l^b = -Le^f \delta_a^{+}\,.
\end{gather}
In particular, we obtain the relation
\begin{gather}
 k_c l^c = KLe^f\,.
\end{gather}

We define the expansions $\Theta_{\pm}$ as
\begin{gather}
 \Theta_{\pm} = \frac{1}{2} s^{ab} \Lie_{\pm} s_{ab} = 2 \frac{\partial_{\pm} r}{r}\,,
\end{gather}
\noindent where we denote $\partial_{\pm} = \delta^c_{\pm} \partial_{c}$.

In terms of the basis vectors, the Christoffel symbols are written as
\begin{gather}
\Gamma_{\alpha \beta}^{\gamma} =  \delta^{\gamma}_{b} \delta^{a}_{\alpha}\nabla_{a} \delta^b_{\beta}
\end{gather}

We obtain the following components for the Christoffel symbols \eqref{eq:metric-coordinates} in terms of the quantities we defined:
\begin{gather}
 \Gamma_{\pm \pm}^{\pm} =\partial_{\pm} f\,; 
 \quad s^{\alpha}_{\mu} s^{\nu}_{\gamma}\Gamma_{\nu \pm}^\mu = \frac{\Theta_{\pm}}{2}s^\alpha_{\gamma}\,; \quad s^{\mu}_{\alpha} s^{\nu}_{\beta}\Gamma_{\mu \nu}^{\pm} = e^{-f} \frac{\Theta_{\pm}}{2} s_{\alpha \beta}\,;
\end{gather}
\noindent
where we used $s^{\mu}_{\nu}$ as a projector to the spatial directions. The other components are null or just correspond to the components of the two-dimensional sphere whose metric is $s_{ab}$.

Computing the space-space components of the Ricci tensor we obtain
\begin{gather}
 R_{\beta \gamma} = \left[ e^{-f} \partial_{(-}\Theta_{+)} + e^{-f} \Theta_+ \Theta_- + \frac{1}{r^2}\right] s_{\beta \gamma}\,.
\end{gather}

The Ricci scalar is
\begin{gather}
 R = g^{a b} R_{a b} = 2g^{+-}R_{+-} + s^{ab}R_{ab} = -2e^{-f}R_{+-} + 2 \left[ e^{-f} \partial_{(-}\Theta_{+)} + e^{-f} \Theta_+ \Theta_- + \frac{1}{r^2}\right]\,.
\end{gather}
The $(+-)$ component of the Einstein tensor is given by
\begin{gather}
 G_{+-} = R_{+-} - \frac{1}{2} g_{+-} R = \nonumber\\ 
 = R_{+-} + e^f\left[- e^{-f}R_{+-}+\left[ e^{-f} \partial_{(-}\Theta_{+)} +  e^{-f} \Theta_+ \Theta_- + \frac{1}{r^2}\right] \right] = \nonumber\\
 =  \partial_{(-} \Theta_{+)}+  \Theta_+ \Theta_- + \frac{e^f}{r^2}
\end{gather}
using the fact that $\partial_- \Theta_+ = \partial_+ \Theta_-$ and Einstein's equations, we finally obtain
\begin{gather}
 \Lie_{\pm} \Theta_{\mp} = - \Theta_+ \Theta_- - \frac{e^f}{r^2} + 8 \pi  T_{+-}\, , \label{Raychauduri2}
\end{gather}

Generalizing the expression \eqref{Raychauduri2} for our null basis $k^a = K\delta^a_{+}$, and $l^a = L\delta^a_{-}$, we find
\begin{gather}
 \Lie_l \Theta_{(k)} = L\left[ \Lie_- (K\Theta_+)\right] = L\left( K \Lie_- \Theta_+ + \Theta_+ \Lie_- K \right) = \nonumber\\
  =\Theta_+ L \Lie_- K + KL \left(- \Theta_+ \Theta_- - \frac{e^f}{r^2} + 8 \pi  T_{+-}\right) = \nonumber\\
  =\frac{\Lie_l K}{K} \Theta_{(k)} + \left( - \Theta_{(k)} \Theta_{(l)} - \frac{k^cl_c}{r^2} + 8\pi  T_{ab}k^al^b\right)\,.
\end{gather}
The last step is identifying the term $\frac{\Lie_l K}{K}$. We can write $K = k^b\delta^{+}_b$, therefore
\begin{gather}
 \frac{\Lie_l K}{K} = \frac{1}{K} l^a \nabla_a (K) = \frac{1}{K}l^a \nabla_a (k^b\delta_b^{+}) = \nonumber\\
 = \frac{1}{K} \left( \delta_b^{+} l^a \nabla_a k^b + l^a k^b \nabla_a \delta^+_b \right) = \nonumber\\
 = \frac{1}{e^f KL} l_b l^a \nabla_a k^b = \frac{1}{k_c l^c} l^a l^b \nabla_a k_b = - \nu_l\,,
\end{gather}
\noindent where we used $l^ak^b \nabla_a \delta^+_b = LK \delta^a_{-}\delta^b_{+} \nabla_{a} \delta^{+}_b = LK\Gamma_{+-}^+ =0$, since this Christoffel symbols vanishes.

Therefore, we find finally
\begin{gather}
 \Lie_l \Theta_{(k)} = -\Theta_{(k)} \Theta_{(l)} - \nu_l \Theta_{(k)} - \frac{k^c l_c}{r^2} + 8 \pi T_{ab}k^al^b\,,
\end{gather}
\noindent or, in the symmetrical form
\begin{gather}
    \Lie_l \Theta_{(k)} +\Lie_k \Theta_{(l)} = -2\Theta_{(k)} \Theta_{(l)} - \nu_l\Theta_{(k)} - \nu_k \Theta_{(l)} - \frac{2k^c l_c}{r^2} +16 \pi T_{ab}k^al^b\,, \label{cross}
\end{gather}

 \bibliography{shortnames,referencias}

\newcommand{\grg}{Gen.\ Relativ.\ Gravit.\@}
\begin{thebibliography}{50}%
\makeatletter
\providecommand \@ifxundefined [1]{%
 \@ifx{#1\undefined}
}%
\providecommand \@ifnum [1]{%
 \ifnum #1\expandafter \@firstoftwo
 \else \expandafter \@secondoftwo
 \fi
}%
\providecommand \@ifx [1]{%
 \ifx #1\expandafter \@firstoftwo
 \else \expandafter \@secondoftwo
 \fi
}%
\providecommand \natexlab [1]{#1}%
\providecommand \enquote  [1]{``#1''}%
\providecommand \bibnamefont  [1]{#1}%
\providecommand \bibfnamefont [1]{#1}%
\providecommand \citenamefont [1]{#1}%
\providecommand \href@noop [0]{\@secondoftwo}%
\providecommand \href [0]{\begingroup \@sanitize@url \@href}%
\providecommand \@href[1]{\@@startlink{#1}\@@href}%
\providecommand \@@href[1]{\endgroup#1\@@endlink}%
\providecommand \@sanitize@url [0]{\catcode `\\12\catcode `\$12\catcode
  `\&12\catcode `\#12\catcode `\^12\catcode `\_12\catcode `\%12\relax}%
\providecommand \@@startlink[1]{}%
\providecommand \@@endlink[0]{}%
\providecommand \url  [0]{\begingroup\@sanitize@url \@url }%
\providecommand \@url [1]{\endgroup\@href {#1}{\urlprefix }}%
\providecommand \urlprefix  [0]{URL }%
\providecommand \Eprint [0]{\href }%
\providecommand \doibase [0]{http://dx.doi.org/}%
\providecommand \selectlanguage [0]{\@gobble}%
\providecommand \bibinfo  [0]{\@secondoftwo}%
\providecommand \bibfield  [0]{\@secondoftwo}%
\providecommand \translation [1]{[#1]}%
\providecommand \BibitemOpen [0]{}%
\providecommand \bibitemStop [0]{}%
\providecommand \bibitemNoStop [0]{.\EOS\space}%
\providecommand \EOS [0]{\spacefactor3000\relax}%
\providecommand \BibitemShut  [1]{\csname bibitem#1\endcsname}%
\let\auto@bib@innerbib\@empty
\bibitem [{\citenamefont {Penrose}(2005)}]{Penrose:2005bg}%
  \BibitemOpen
  \bibfield  {author} {\bibinfo {author} {\bibfnamefont {R.}~\bibnamefont
  {Penrose}},\ }\href@noop {} {\emph {\bibinfo {title} {{A complete guide to
  the laws of the universe}}}}\ (\bibinfo  {publisher} {Jonathan Cape},\
  \bibinfo {address} {London},\ \bibinfo {year} {2005})\BibitemShut {NoStop}%
\bibitem [{\citenamefont {d'Inverno}(1992)}]{dInverno:1992gxs}%
  \BibitemOpen
  \bibfield  {author} {\bibinfo {author} {\bibfnamefont {R.}~\bibnamefont
  {d'Inverno}},\ }\href@noop {} {\emph {\bibinfo {title} {{Introducing
  Einstein's relativity}}}}\ (\bibinfo {year} {1992})\BibitemShut {NoStop}%
\bibitem [{\citenamefont {Weinberg}(1972)}]{weinberg}%
  \BibitemOpen
  \bibfield  {author} {\bibinfo {author} {\bibfnamefont {S.}~\bibnamefont
  {Weinberg}},\ }\href@noop {} {\emph {\bibinfo {title} {Gravitation and
  Cosmology}}}\ (\bibinfo  {publisher} {John Wiley \& Sons},\ \bibinfo {year}
  {1972})\BibitemShut {NoStop}%
\bibitem [{\citenamefont {{Kopczynski}}\ and\ \citenamefont
  {{Trautman}}(1992)}]{Kopczynski-Trautman-book}%
  \BibitemOpen
  \bibfield  {author} {\bibinfo {author} {\bibfnamefont {W.}~\bibnamefont
  {{Kopczynski}}}\ and\ \bibinfo {author} {\bibfnamefont {A.}~\bibnamefont
  {{Trautman}}},\ }\href@noop {} {\emph {\bibinfo {title} {``Spacetime and
  gravitation"}}}\ (\bibinfo {year} {1992})\BibitemShut {NoStop}%
\bibitem [{\citenamefont {Hawking}\ and\ \citenamefont
  {Ellis}(1973)}]{hawking}%
  \BibitemOpen
  \bibfield  {author} {\bibinfo {author} {\bibfnamefont {S.~W.}\ \bibnamefont
  {Hawking}}\ and\ \bibinfo {author} {\bibfnamefont {G.~F.~R.}\ \bibnamefont
  {Ellis}},\ }\href@noop {} {\emph {\bibinfo {title} {The Large Scale Structure
  of Space-Time}}}\ (\bibinfo  {publisher} {Cambridge University Press},\
  \bibinfo {address} {Cambridge},\ \bibinfo {year} {1973})\BibitemShut
  {NoStop}%
\bibitem [{\citenamefont {Hayward}(1994)}]{Hayward:1993mw}%
  \BibitemOpen
  \bibfield  {author} {\bibinfo {author} {\bibfnamefont {S.~A.}\ \bibnamefont
  {Hayward}},\ }\href {\doibase 10.1103/PhysRevD.49.6467} {\bibfield  {journal}
  {\bibinfo  {journal} {Phys.\ Rev.\ D}\ }\textbf {\bibinfo {volume} {49}},\
  \bibinfo {pages} {6467} (\bibinfo {year} {1994})},\ \Eprint
  {http://arxiv.org/abs/gr-qc/9303006} {arXiv:gr-qc/9303006} \BibitemShut
  {NoStop}%
\bibitem [{\citenamefont {Bak}\ and\ \citenamefont {Rey}(2000)}]{Bak:1999hd}%
  \BibitemOpen
  \bibfield  {author} {\bibinfo {author} {\bibfnamefont {D.}~\bibnamefont
  {Bak}}\ and\ \bibinfo {author} {\bibfnamefont {S.-J.}\ \bibnamefont {Rey}},\
  }\href {\doibase 10.1088/0264-9381/17/15/101} {\bibfield  {journal} {\bibinfo
   {journal} {Class. Quant. Grav.}\ }\textbf {\bibinfo {volume} {17}},\
  \bibinfo {pages} {L83} (\bibinfo {year} {2000})},\ \Eprint
  {http://arxiv.org/abs/hep-th/9902173} {arXiv:hep-th/9902173 [hep-th]}
  \BibitemShut {NoStop}%
\bibitem [{\citenamefont {Cai}\ and\ \citenamefont {Cao}(2007)}]{Cai:2006rs}%
  \BibitemOpen
  \bibfield  {author} {\bibinfo {author} {\bibfnamefont {R.-G.}\ \bibnamefont
  {Cai}}\ and\ \bibinfo {author} {\bibfnamefont {L.-M.}\ \bibnamefont {Cao}},\
  }\href {\doibase 10.1103/PhysRevD.75.064008} {\bibfield  {journal} {\bibinfo
  {journal} {Phys. Rev.}\ }\textbf {\bibinfo {volume} {D75}},\ \bibinfo {pages}
  {064008} (\bibinfo {year} {2007})},\ \Eprint
  {http://arxiv.org/abs/gr-qc/0611071} {arXiv:gr-qc/0611071 [gr-qc]}
  \BibitemShut {NoStop}%
\bibitem [{\citenamefont {Hayward}(1998)}]{Hayward:1997jp}%
  \BibitemOpen
  \bibfield  {author} {\bibinfo {author} {\bibfnamefont {S.~A.}\ \bibnamefont
  {Hayward}},\ }\href {\doibase 10.1088/0264-9381/15/10/017} {\bibfield
  {journal} {\bibinfo  {journal} {Class. Quant. Grav.}\ }\textbf {\bibinfo
  {volume} {15}},\ \bibinfo {pages} {3147} (\bibinfo {year} {1998})},\ \Eprint
  {http://arxiv.org/abs/gr-qc/9710089} {arXiv:gr-qc/9710089 [gr-qc]}
  \BibitemShut {NoStop}%
\bibitem [{\citenamefont {Binetruy}\ and\ \citenamefont
  {Helou}(2015)}]{Binetruy:2014ela}%
  \BibitemOpen
  \bibfield  {author} {\bibinfo {author} {\bibfnamefont {P.}~\bibnamefont
  {Binetruy}}\ and\ \bibinfo {author} {\bibfnamefont {A.}~\bibnamefont
  {Helou}},\ }\href {\doibase 10.1088/0264-9381/32/20/205006} {\bibfield
  {journal} {\bibinfo  {journal} {Class. Quant. Grav.}\ }\textbf {\bibinfo
  {volume} {32}},\ \bibinfo {pages} {205006} (\bibinfo {year} {2015})},\
  \Eprint {http://arxiv.org/abs/1406.1658} {arXiv:1406.1658 [gr-qc]}
  \BibitemShut {NoStop}%
\bibitem [{\citenamefont {Sachs}(1964)}]{Sachs:1964zza}%
  \BibitemOpen
  \bibfield  {author} {\bibinfo {author} {\bibfnamefont {R.~K.}\ \bibnamefont
  {Sachs}},\ }in\ \href@noop {} {\emph {\bibinfo {booktitle} {{Les Houches
  Summer Shcool of Theoretical Physics}: {Relativity, Groups and Topology}}}}\
  (\bibinfo {year} {1964})\ pp.\ \bibinfo {pages} {523--564}\BibitemShut
  {NoStop}%
\bibitem [{\citenamefont {Newman}\ and\ \citenamefont
  {Penrose}(1962)}]{Newman:1961qr}%
  \BibitemOpen
  \bibfield  {author} {\bibinfo {author} {\bibfnamefont {E.}~\bibnamefont
  {Newman}}\ and\ \bibinfo {author} {\bibfnamefont {R.}~\bibnamefont
  {Penrose}},\ }\href {\doibase 10.1063/1.1724257} {\bibfield  {journal}
  {\bibinfo  {journal} {J. Math. Phys.}\ }\textbf {\bibinfo {volume} {3}},\
  \bibinfo {pages} {566} (\bibinfo {year} {1962})}\BibitemShut {NoStop}%
\bibitem [{\citenamefont {Chandrasekhar}(1980)}]{Chandrasekhar:1980gb}%
  \BibitemOpen
  \bibfield  {author} {\bibinfo {author} {\bibfnamefont {S.}~\bibnamefont
  {Chandrasekhar}},\ }\enquote {\bibinfo {title} {{AN INTRODUCTION TO THE
  THEORY OF THE KERR METRIC AND ITS PERTURBATIONS}},}\ in\ \href@noop {} {\emph
  {\bibinfo {booktitle} {{General Relativity}: {An Einstein Centenary
  Survey}}}}\ (\bibinfo {year} {1980})\ pp.\ \bibinfo {pages}
  {370--453}\BibitemShut {NoStop}%
\bibitem [{\citenamefont {d'Inverno}\ and\ \citenamefont
  {Smallwood}(1980)}]{dInverno:1980kaa}%
  \BibitemOpen
  \bibfield  {author} {\bibinfo {author} {\bibfnamefont {R.~A.}\ \bibnamefont
  {d'Inverno}}\ and\ \bibinfo {author} {\bibfnamefont {J.}~\bibnamefont
  {Smallwood}},\ }\href {\doibase 10.1103/PhysRevD.22.1233} {\bibfield
  {journal} {\bibinfo  {journal} {Phys. Rev. D}\ }\textbf {\bibinfo {volume}
  {22}},\ \bibinfo {pages} {1233} (\bibinfo {year} {1980})}\BibitemShut
  {NoStop}%
\bibitem [{\citenamefont {Brady}\ \emph {et~al.}(1996)\citenamefont {Brady},
  \citenamefont {Droz}, \citenamefont {Israel},\ and\ \citenamefont
  {Morsink}}]{Brady:1995na}%
  \BibitemOpen
  \bibfield  {author} {\bibinfo {author} {\bibfnamefont {P.~R.}\ \bibnamefont
  {Brady}}, \bibinfo {author} {\bibfnamefont {S.}~\bibnamefont {Droz}},
  \bibinfo {author} {\bibfnamefont {W.}~\bibnamefont {Israel}}, \ and\ \bibinfo
  {author} {\bibfnamefont {S.~M.}\ \bibnamefont {Morsink}},\ }\href {\doibase
  10.1088/0264-9381/13/8/015} {\bibfield  {journal} {\bibinfo  {journal}
  {Class. Quant. Grav.}\ }\textbf {\bibinfo {volume} {13}},\ \bibinfo {pages}
  {2211} (\bibinfo {year} {1996})},\ \Eprint
  {http://arxiv.org/abs/gr-qc/9510040} {arXiv:gr-qc/9510040} \BibitemShut
  {NoStop}%
\bibitem [{\citenamefont {Fleury}\ \emph {et~al.}(2016)\citenamefont {Fleury},
  \citenamefont {Nugier},\ and\ \citenamefont {Fanizza}}]{Fleury:2016htl}%
  \BibitemOpen
  \bibfield  {author} {\bibinfo {author} {\bibfnamefont {P.}~\bibnamefont
  {Fleury}}, \bibinfo {author} {\bibfnamefont {F.}~\bibnamefont {Nugier}}, \
  and\ \bibinfo {author} {\bibfnamefont {G.}~\bibnamefont {Fanizza}},\ }\href
  {\doibase 10.1088/1475-7516/2016/06/008} {\bibfield  {journal} {\bibinfo
  {journal} {JCAP}\ }\textbf {\bibinfo {volume} {06}},\ \bibinfo {pages} {008}
  (\bibinfo {year} {2016})},\ \Eprint {http://arxiv.org/abs/1602.04461}
  {arXiv:1602.04461 [gr-qc]} \BibitemShut {NoStop}%
\bibitem [{\citenamefont {Nugier}(2016)}]{Nugier:2016ebh}%
  \BibitemOpen
  \bibfield  {author} {\bibinfo {author} {\bibfnamefont {F.}~\bibnamefont
  {Nugier}},\ }\href {\doibase 10.1088/1475-7516/2016/09/019} {\bibfield
  {journal} {\bibinfo  {journal} {JCAP}\ }\textbf {\bibinfo {volume} {09}},\
  \bibinfo {pages} {019} (\bibinfo {year} {2016})},\ \Eprint
  {http://arxiv.org/abs/1606.08296} {arXiv:1606.08296 [gr-qc]} \BibitemShut
  {NoStop}%
\bibitem [{\citenamefont {Nakonieczna}\ \emph {et~al.}(2018)\citenamefont
  {Nakonieczna}, \citenamefont {Nakonieczny},\ and\ \citenamefont
  {Yeom}}]{Nakonieczna:2018tih}%
  \BibitemOpen
  \bibfield  {author} {\bibinfo {author} {\bibfnamefont {A.}~\bibnamefont
  {Nakonieczna}}, \bibinfo {author} {\bibfnamefont {L.}~\bibnamefont
  {Nakonieczny}}, \ and\ \bibinfo {author} {\bibfnamefont {D.-h.}\ \bibnamefont
  {Yeom}},\ }\href {\doibase 10.1142/S0218271819300064} {\bibfield  {journal}
  {\bibinfo  {journal} {Int. J. Mod. Phys. D}\ }\textbf {\bibinfo {volume}
  {28}},\ \bibinfo {pages} {1930006} (\bibinfo {year} {2018})},\ \Eprint
  {http://arxiv.org/abs/1805.12362} {arXiv:1805.12362 [gr-qc]} \BibitemShut
  {NoStop}%
\bibitem [{\citenamefont {Maciel}\ \emph {et~al.}(2015)\citenamefont {Maciel},
  \citenamefont {Le~Delliou},\ and\ \citenamefont {Mimoso}}]{Maciel:2015vva}%
  \BibitemOpen
  \bibfield  {author} {\bibinfo {author} {\bibfnamefont {A.}~\bibnamefont
  {Maciel}}, \bibinfo {author} {\bibfnamefont {M.}~\bibnamefont {Le~Delliou}},
  \ and\ \bibinfo {author} {\bibfnamefont {J.~P.}\ \bibnamefont {Mimoso}},\
  }\href {\doibase 10.1103/PhysRevD.92.083525} {\bibfield  {journal} {\bibinfo
  {journal} {Phys.\ Rev.\ D}\ }\textbf {\bibinfo {volume} {D92}},\ \bibinfo
  {pages} {083525} (\bibinfo {year} {2015})},\ \Eprint
  {http://arxiv.org/abs/1506.07122} {arXiv:1506.07122 [gr-qc]} \BibitemShut
  {NoStop}%
\bibitem [{\citenamefont {Maciel}\ \emph {et~al.}(2018)\citenamefont {Maciel},
  \citenamefont {Le~Delliou},\ and\ \citenamefont {Mimoso}}]{Maciel:2018tnc}%
  \BibitemOpen
  \bibfield  {author} {\bibinfo {author} {\bibfnamefont {A.}~\bibnamefont
  {Maciel}}, \bibinfo {author} {\bibfnamefont {M.}~\bibnamefont {Le~Delliou}},
  \ and\ \bibinfo {author} {\bibfnamefont {J.~P.}\ \bibnamefont {Mimoso}},\
  }\href {\doibase 10.1103/PhysRevD.98.024016} {\bibfield  {journal} {\bibinfo
  {journal} {Phys. Rev.}\ }\textbf {\bibinfo {volume} {D98}},\ \bibinfo {pages}
  {024016} (\bibinfo {year} {2018})},\ \Eprint
  {http://arxiv.org/abs/1803.11547} {arXiv:1803.11547 [gr-qc]} \BibitemShut
  {NoStop}%
\bibitem [{\citenamefont {Maciel}\ \emph
  {et~al.}(2020{\natexlab{a}})\citenamefont {Maciel}, \citenamefont
  {Le~Delliou},\ and\ \citenamefont {Mimoso}}]{Maciel:2019grc}%
  \BibitemOpen
  \bibfield  {author} {\bibinfo {author} {\bibfnamefont {A.}~\bibnamefont
  {Maciel}}, \bibinfo {author} {\bibfnamefont {M.}~\bibnamefont {Le~Delliou}},
  \ and\ \bibinfo {author} {\bibfnamefont {J.~P.}\ \bibnamefont {Mimoso}},\
  }\href {\doibase 10.1088/1361-6382/ab8759} {\bibfield  {journal} {\bibinfo
  {journal} {Class. Quant. Grav.}\ }\textbf {\bibinfo {volume} {37}},\ \bibinfo
  {pages} {125005} (\bibinfo {year} {2020}{\natexlab{a}})},\ \Eprint
  {http://arxiv.org/abs/1910.13225} {arXiv:1910.13225 [gr-qc]} \BibitemShut
  {NoStop}%
\bibitem [{\citenamefont {Clarkson}\ and\ \citenamefont
  {Barrett}(2003)}]{Clarkson:2002jz}%
  \BibitemOpen
  \bibfield  {author} {\bibinfo {author} {\bibfnamefont {C.~A.}\ \bibnamefont
  {Clarkson}}\ and\ \bibinfo {author} {\bibfnamefont {R.~K.}\ \bibnamefont
  {Barrett}},\ }\href {\doibase 10.1088/0264-9381/20/18/301} {\bibfield
  {journal} {\bibinfo  {journal} {Classical Quantum Gravity}\ }\textbf
  {\bibinfo {volume} {20}},\ \bibinfo {pages} {3855} (\bibinfo {year}
  {2003})},\ \Eprint {http://arxiv.org/abs/gr-qc/0209051} {arXiv:gr-qc/0209051
  [gr-qc]} \BibitemShut {NoStop}%
\bibitem [{\citenamefont {Clarkson}(2007)}]{Clarkson:2007yp}%
  \BibitemOpen
  \bibfield  {author} {\bibinfo {author} {\bibfnamefont {C.~A.}\ \bibnamefont
  {Clarkson}},\ }\href {\doibase 10.1103/PhysRevD.76.104034} {\bibfield
  {journal} {\bibinfo  {journal} {Phys.\ Rev.\ D}\ }\textbf {\bibinfo {volume}
  {76}},\ \bibinfo {pages} {104034} (\bibinfo {year} {2007})},\ \Eprint
  {http://arxiv.org/abs/0708.1398} {arXiv:0708.1398 [gr-qc]} \BibitemShut
  {NoStop}%
\bibitem [{\citenamefont {Goswami}\ and\ \citenamefont
  {Ellis}(2011)}]{Goswami:2011ft}%
  \BibitemOpen
  \bibfield  {author} {\bibinfo {author} {\bibfnamefont {R.}~\bibnamefont
  {Goswami}}\ and\ \bibinfo {author} {\bibfnamefont {G.~F.~R.}\ \bibnamefont
  {Ellis}},\ }\href {\doibase 10.1007/s10714-011-1172-z} {\bibfield  {journal}
  {\bibinfo  {journal} {Gen. Rel. Grav.}\ }\textbf {\bibinfo {volume} {43}},\
  \bibinfo {pages} {2157} (\bibinfo {year} {2011})},\ \Eprint
  {http://arxiv.org/abs/1101.4520} {arXiv:1101.4520 [gr-qc]} \BibitemShut
  {NoStop}%
\bibitem [{\citenamefont {Goswami}\ and\ \citenamefont
  {Ellis}(2012)}]{Goswami:2012jf}%
  \BibitemOpen
  \bibfield  {author} {\bibinfo {author} {\bibfnamefont {R.}~\bibnamefont
  {Goswami}}\ and\ \bibinfo {author} {\bibfnamefont {G.~F.~R.}\ \bibnamefont
  {Ellis}},\ }\href {\doibase 10.1007/s10714-012-1376-x} {\bibfield  {journal}
  {\bibinfo  {journal} {Gen. Rel. Grav.}\ }\textbf {\bibinfo {volume} {44}},\
  \bibinfo {pages} {2037} (\bibinfo {year} {2012})},\ \Eprint
  {http://arxiv.org/abs/1202.0240} {arXiv:1202.0240 [gr-qc]} \BibitemShut
  {NoStop}%
\bibitem [{\citenamefont {Ellis}\ and\ \citenamefont
  {Goswami}(2013)}]{Ellis:2013dla}%
  \BibitemOpen
  \bibfield  {author} {\bibinfo {author} {\bibfnamefont {G.~F.~R.}\
  \bibnamefont {Ellis}}\ and\ \bibinfo {author} {\bibfnamefont
  {R.}~\bibnamefont {Goswami}},\ }\bibfield  {booktitle} {\emph {\bibinfo
  {booktitle} {{Progress in Mathematical Relativity, Gravitation and Cosmology:
  Proceedings, Spanish Relativity Meeting ERE2012, University of Minho,
  Guimarães, Portugal, September 3-7, 2012}}},\ }\href {\doibase
  10.1007/s10714-013-1568-z} {\bibfield  {journal} {\bibinfo  {journal} {Gen.
  Rel. Grav.}\ }\textbf {\bibinfo {volume} {45}},\ \bibinfo {pages} {2123}
  (\bibinfo {year} {2013})},\ \Eprint {http://arxiv.org/abs/1304.3253}
  {arXiv:1304.3253 [gr-qc]} \BibitemShut {NoStop}%
\bibitem [{\citenamefont {G\l{}\'od}(2020)}]{Glod:2020adq}%
  \BibitemOpen
  \bibfield  {author} {\bibinfo {author} {\bibfnamefont {K.}~\bibnamefont
  {G\l{}\'od}},\ }\href {\doibase 10.1103/PhysRevD.101.024021} {\bibfield
  {journal} {\bibinfo  {journal} {Phys. Rev. D}\ }\textbf {\bibinfo {volume}
  {101}},\ \bibinfo {pages} {024021} (\bibinfo {year} {2020})},\ \Eprint
  {http://arxiv.org/abs/2001.02782} {arXiv:2001.02782 [gr-qc]} \BibitemShut
  {NoStop}%
\bibitem [{\citenamefont {Maciel}(2016)}]{Maciel:2015ypv}%
  \BibitemOpen
  \bibfield  {author} {\bibinfo {author} {\bibfnamefont {A.}~\bibnamefont
  {Maciel}},\ }\href {\doibase 10.1103/PhysRevD.93.104013} {\bibfield
  {journal} {\bibinfo  {journal} {Phys. Rev.}\ }\textbf {\bibinfo {volume}
  {D93}},\ \bibinfo {pages} {104013} (\bibinfo {year} {2016})},\ \Eprint
  {http://arxiv.org/abs/1511.08663} {arXiv:1511.08663 [gr-qc]} \BibitemShut
  {NoStop}%
\bibitem [{\citenamefont {{Painlev{\'e}}}(1921)}]{Painleve1921}%
  \BibitemOpen
  \bibfield  {author} {\bibinfo {author} {\bibfnamefont {P.}~\bibnamefont
  {{Painlev{\'e}}}},\ }\href@noop {} {\bibfield  {journal} {\bibinfo  {journal}
  {Comptes Rendus Academie des Sciences (serie non specifiee)}\ }\textbf
  {\bibinfo {volume} {173}},\ \bibinfo {pages} {677} (\bibinfo {year}
  {1921})}\BibitemShut {NoStop}%
\bibitem [{\citenamefont {Gullstrand}(1922)}]{gullstrand-1922}%
  \BibitemOpen
  \bibfield  {author} {\bibinfo {author} {\bibfnamefont {A.}~\bibnamefont
  {Gullstrand}},\ }\href@noop {} {\bibfield  {journal} {\bibinfo  {journal}
  {Ark.\ Mat.\ Astron.\ Fys.\@}\ }\textbf {\bibinfo {volume} {16}},\ \bibinfo
  {pages} {1} (\bibinfo {year} {1922})}\BibitemShut {NoStop}%
\bibitem [{\citenamefont {Gautreau}(1984)}]{Gautreau:1984PhRvD186}%
  \BibitemOpen
  \bibfield  {author} {\bibinfo {author} {\bibfnamefont {R.}~\bibnamefont
  {Gautreau}},\ }\href {\doibase 10.1103/PhysRevD.29.186} {\bibfield  {journal}
  {\bibinfo  {journal} {Phys. Rev. D}\ }\textbf {\bibinfo {volume} {29}},\
  \bibinfo {pages} {186} (\bibinfo {year} {1984})}\BibitemShut {NoStop}%
\bibitem [{\citenamefont {Lasky}\ and\ \citenamefont
  {Lun}(2006)}]{LaskyLun06b}%
  \BibitemOpen
  \bibfield  {author} {\bibinfo {author} {\bibfnamefont {P.~D.}\ \bibnamefont
  {Lasky}}\ and\ \bibinfo {author} {\bibfnamefont {A.~W.~C.}\ \bibnamefont
  {Lun}},\ }\href {\doibase 10.1103/PhysRevD.74.084013} {\bibfield  {journal}
  {\bibinfo  {journal} {Phys.\ Rev.\ D}\ }\textbf {\bibinfo {volume} {74}},\
  \bibinfo {pages} {084013} (\bibinfo {year} {2006})},\ \Eprint
  {http://arxiv.org/abs/gr-qc/0606055} {arXiv:gr-qc/0606055} \BibitemShut
  {NoStop}%
\bibitem [{\citenamefont {Lasky}\ and\ \citenamefont
  {Lun}(2007{\natexlab{a}})}]{LaskyLun07}%
  \BibitemOpen
  \bibfield  {author} {\bibinfo {author} {\bibfnamefont {P.~D.}\ \bibnamefont
  {Lasky}}\ and\ \bibinfo {author} {\bibfnamefont {A.~W.~C.}\ \bibnamefont
  {Lun}},\ }\href {\doibase 10.1103/PhysRevD.75.104010} {\bibfield  {journal}
  {\bibinfo  {journal} {Phys.\ Rev.\ D}\ }\textbf {\bibinfo {volume} {75}},\
  \bibinfo {pages} {104010} (\bibinfo {year} {2007}{\natexlab{a}})},\ \Eprint
  {http://arxiv.org/abs/0704.3634} {arXiv:0704.3634 [gr-qc]} \BibitemShut
  {NoStop}%
\bibitem [{\citenamefont {Faraoni}\ and\ \citenamefont
  {Vachon}(2020)}]{Faraoni:2020ehi}%
  \BibitemOpen
  \bibfield  {author} {\bibinfo {author} {\bibfnamefont {V.}~\bibnamefont
  {Faraoni}}\ and\ \bibinfo {author} {\bibfnamefont {G.}~\bibnamefont
  {Vachon}},\ }\href {\doibase 10.1140/epjc/s10052-020-8345-4} {\bibfield
  {journal} {\bibinfo  {journal} {Eur. Phys. J. C}\ }\textbf {\bibinfo {volume}
  {80}},\ \bibinfo {pages} {771} (\bibinfo {year} {2020})},\ \Eprint
  {http://arxiv.org/abs/2006.10827} {arXiv:2006.10827 [gr-qc]} \BibitemShut
  {NoStop}%
\bibitem [{\citenamefont {Sussman}(2009)}]{Sussman:2008wx}%
  \BibitemOpen
  \bibfield  {author} {\bibinfo {author} {\bibfnamefont {R.~A.}\ \bibnamefont
  {Sussman}},\ }\href {\doibase 10.1103/PhysRevD.79.025009} {\bibfield
  {journal} {\bibinfo  {journal} {Phys. Rev. D}\ }\textbf {\bibinfo {volume}
  {79}},\ \bibinfo {pages} {025009} (\bibinfo {year} {2009})},\ \Eprint
  {http://arxiv.org/abs/0801.3324} {arXiv:0801.3324 [gr-qc]} \BibitemShut
  {NoStop}%
\bibitem [{\citenamefont {Ellis}(2009)}]{Ellis:1971pg}%
  \BibitemOpen
  \bibfield  {author} {\bibinfo {author} {\bibfnamefont {G.~F.~R.}\
  \bibnamefont {Ellis}},\ }\href {\doibase 10.1007/s10714-009-0760-7}
  {\bibfield  {journal} {\bibinfo  {journal} {\grg}\ }\textbf {\bibinfo
  {volume} {41}},\ \bibinfo {pages} {581} (\bibinfo {year} {2009})}\BibitemShut
  {NoStop}%
\bibitem [{\citenamefont {Codur}\ and\ \citenamefont
  {Marinoni}(2021)}]{Codur:2021wrt}%
  \BibitemOpen
  \bibfield  {author} {\bibinfo {author} {\bibfnamefont {R.}~\bibnamefont
  {Codur}}\ and\ \bibinfo {author} {\bibfnamefont {C.}~\bibnamefont
  {Marinoni}},\ }\href {\doibase 10.1103/PhysRevD.104.123531} {\bibfield
  {journal} {\bibinfo  {journal} {Phys. Rev. D}\ }\textbf {\bibinfo {volume}
  {104}},\ \bibinfo {pages} {123531} (\bibinfo {year} {2021})},\ \Eprint
  {http://arxiv.org/abs/2107.04868} {arXiv:2107.04868 [gr-qc]} \BibitemShut
  {NoStop}%
\bibitem [{\citenamefont {Hayward}(1993)}]{hayward-1993}%
  \BibitemOpen
  \bibfield  {author} {\bibinfo {author} {\bibfnamefont {S.~A.}\ \bibnamefont
  {Hayward}},\ }\href {\doibase 10.1088/0264-9381/10/4/013} {\bibfield
  {journal} {\bibinfo  {journal} {Classical Quantum Gravity}\ }\textbf
  {\bibinfo {volume} {10}},\ \bibinfo {pages} {779} (\bibinfo {year}
  {1993})}\BibitemShut {NoStop}%
\bibitem [{\citenamefont {Herrera}\ \emph {et~al.}(2010)\citenamefont
  {Herrera}, \citenamefont {{Di Prisco}}, \citenamefont {Ospino},\ and\
  \citenamefont {Carot}}]{herrera-2010a}%
  \BibitemOpen
  \bibfield  {author} {\bibinfo {author} {\bibfnamefont {L.}~\bibnamefont
  {Herrera}}, \bibinfo {author} {\bibfnamefont {A.}~\bibnamefont {{Di
  Prisco}}}, \bibinfo {author} {\bibfnamefont {J.}~\bibnamefont {Ospino}}, \
  and\ \bibinfo {author} {\bibfnamefont {J.}~\bibnamefont {Carot}},\ }\href
  {\doibase 10.1103/PhysRevD.82.024021} {\bibfield  {journal} {\bibinfo
  {journal} {Phys.\ Rev.\ D}\ }\textbf {\bibinfo {volume} {82}},\ \bibinfo
  {pages} {024021} (\bibinfo {year} {2010})},\ \Eprint
  {http://arxiv.org/abs/1006.2149} {arXiv:1006.2149 [gr-qc]} \BibitemShut
  {NoStop}%
\bibitem [{\citenamefont {Herrera}\ \emph {et~al.}(2021)\citenamefont
  {Herrera}, \citenamefont {Prisco},\ and\ \citenamefont
  {Ospino}}]{Herrera:2021wpe}%
  \BibitemOpen
  \bibfield  {author} {\bibinfo {author} {\bibfnamefont {L.}~\bibnamefont
  {Herrera}}, \bibinfo {author} {\bibfnamefont {A.~D.}\ \bibnamefont {Prisco}},
  \ and\ \bibinfo {author} {\bibfnamefont {J.}~\bibnamefont {Ospino}},\ }\href
  {\doibase 10.3390/e23091219} {\bibfield  {journal} {\bibinfo  {journal}
  {Entropy}\ }\textbf {\bibinfo {volume} {23}},\ \bibinfo {pages} {1219}
  (\bibinfo {year} {2021})},\ \Eprint {http://arxiv.org/abs/2110.01888}
  {arXiv:2110.01888 [gr-qc]} \BibitemShut {NoStop}%
\bibitem [{\citenamefont {Carot}\ \emph {et~al.}(1999)\citenamefont {Carot},
  \citenamefont {Senovilla},\ and\ \citenamefont {Vera}}]{Carot:1999zm}%
  \BibitemOpen
  \bibfield  {author} {\bibinfo {author} {\bibfnamefont {J.}~\bibnamefont
  {Carot}}, \bibinfo {author} {\bibfnamefont {J.~M.~M.}\ \bibnamefont
  {Senovilla}}, \ and\ \bibinfo {author} {\bibfnamefont {R.}~\bibnamefont
  {Vera}},\ }\href {\doibase 10.1088/0264-9381/16/9/318} {\bibfield  {journal}
  {\bibinfo  {journal} {Class. Quant. Grav.}\ }\textbf {\bibinfo {volume}
  {16}},\ \bibinfo {pages} {3025} (\bibinfo {year} {1999})},\ \Eprint
  {http://arxiv.org/abs/gr-qc/9905059} {arXiv:gr-qc/9905059} \BibitemShut
  {NoStop}%
\bibitem [{\citenamefont {Faraoni}\ and\ \citenamefont
  {C\^ot\'e}(2019)}]{Faraoni:2018fil}%
  \BibitemOpen
  \bibfield  {author} {\bibinfo {author} {\bibfnamefont {V.}~\bibnamefont
  {Faraoni}}\ and\ \bibinfo {author} {\bibfnamefont {J.}~\bibnamefont
  {C\^ot\'e}},\ }\href {\doibase 10.1140/epjc/s10052-019-6829-x} {\bibfield
  {journal} {\bibinfo  {journal} {Eur. Phys. J. C}\ }\textbf {\bibinfo {volume}
  {79}},\ \bibinfo {pages} {318} (\bibinfo {year} {2019})},\ \Eprint
  {http://arxiv.org/abs/1812.06457} {arXiv:1812.06457 [gr-qc]} \BibitemShut
  {NoStop}%
\bibitem [{\citenamefont {Ellis}\ \emph {et~al.}(2012)\citenamefont {Ellis},
  \citenamefont {Maartens},\ and\ \citenamefont
  {MacCallum}}]{EllisMaartensMacCallum2012}%
  \BibitemOpen
  \bibfield  {author} {\bibinfo {author} {\bibfnamefont {G.~F.~R.}\
  \bibnamefont {Ellis}}, \bibinfo {author} {\bibfnamefont {R.}~\bibnamefont
  {Maartens}}, \ and\ \bibinfo {author} {\bibfnamefont {M.~A.~H.}\ \bibnamefont
  {MacCallum}},\ }\href@noop {} {\emph {\bibinfo {title} {{Relativistic
  Cosmology}}}}\ (\bibinfo  {publisher} {Cambridge University Press, 2012},\
  \bibinfo {address} {Cambridge, UK},\ \bibinfo {year} {2012})\BibitemShut
  {NoStop}%
\bibitem [{\citenamefont {Misner}\ and\ \citenamefont
  {Sharp}(1964)}]{MisnerSharp}%
  \BibitemOpen
  \bibfield  {author} {\bibinfo {author} {\bibfnamefont {C.~W.}\ \bibnamefont
  {Misner}}\ and\ \bibinfo {author} {\bibfnamefont {D.~H.}\ \bibnamefont
  {Sharp}},\ }\href {\doibase 10.1103/PhysRev.136.B571} {\bibfield  {journal}
  {\bibinfo  {journal} {Phys.\ Rev.\@}\ }\textbf {\bibinfo {volume} {136}},\
  \bibinfo {pages} {B571} (\bibinfo {year} {1964})}\BibitemShut {NoStop}%
\bibitem [{\citenamefont {Arnowitt}\ \emph {et~al.}(1959)\citenamefont
  {Arnowitt}, \citenamefont {Deser},\ and\ \citenamefont
  {Misner}}]{Arnowitt:1959ah}%
  \BibitemOpen
  \bibfield  {author} {\bibinfo {author} {\bibfnamefont {R.~L.}\ \bibnamefont
  {Arnowitt}}, \bibinfo {author} {\bibfnamefont {S.}~\bibnamefont {Deser}}, \
  and\ \bibinfo {author} {\bibfnamefont {C.~W.}\ \bibnamefont {Misner}},\
  }\href {\doibase 10.1103/PhysRev.116.1322} {\bibfield  {journal} {\bibinfo
  {journal} {Phys.\ Rev.\@}\ }\textbf {\bibinfo {volume} {116}},\ \bibinfo
  {pages} {1322} (\bibinfo {year} {1959})}\BibitemShut {NoStop}%
\bibitem [{\citenamefont {Lasky}\ and\ \citenamefont
  {Lun}(2007{\natexlab{b}})}]{Lasky:2006zz}%
  \BibitemOpen
  \bibfield  {author} {\bibinfo {author} {\bibfnamefont {P.~D.}\ \bibnamefont
  {Lasky}}\ and\ \bibinfo {author} {\bibfnamefont {A.~W.~C.}\ \bibnamefont
  {Lun}},\ }\href {\doibase 10.1103/PhysRevD.75.024031} {\bibfield  {journal}
  {\bibinfo  {journal} {Phys.\ Rev.\ D}\ }\textbf {\bibinfo {volume} {75}},\
  \bibinfo {pages} {024031} (\bibinfo {year} {2007}{\natexlab{b}})},\ \Eprint
  {http://arxiv.org/abs/gr-qc/0612007} {arXiv:gr-qc/0612007 [gr-qc]}
  \BibitemShut {NoStop}%
\bibitem [{\citenamefont {Adler}\ \emph {et~al.}(2005)\citenamefont {Adler},
  \citenamefont {Bjorken}, \citenamefont {Chen},\ and\ \citenamefont
  {Liu}}]{adler-2005}%
  \BibitemOpen
  \bibfield  {author} {\bibinfo {author} {\bibfnamefont {R.~J.}\ \bibnamefont
  {Adler}}, \bibinfo {author} {\bibfnamefont {J.~D.}\ \bibnamefont {Bjorken}},
  \bibinfo {author} {\bibfnamefont {P.}~\bibnamefont {Chen}}, \ and\ \bibinfo
  {author} {\bibfnamefont {J.~S.}\ \bibnamefont {Liu}},\ }\href {\doibase
  10.1119/1.2117187} {\bibfield  {journal} {\bibinfo  {journal} {Am.\ J.\
  Phys.\@}\ }\textbf {\bibinfo {volume} {73}},\ \bibinfo {pages} {1148}
  (\bibinfo {year} {2005})},\ \Eprint {http://arxiv.org/abs/arXiv
  gr-qc/0502040} {arXiv gr-qc/0502040} \BibitemShut {NoStop}%
\bibitem [{\citenamefont {Mimoso}\ \emph {et~al.}(2013)\citenamefont {Mimoso},
  \citenamefont {{Le Delliou}},\ and\ \citenamefont {Mena}}]{Mimoso:2013iga}%
  \BibitemOpen
  \bibfield  {author} {\bibinfo {author} {\bibfnamefont {J.~P.}\ \bibnamefont
  {Mimoso}}, \bibinfo {author} {\bibfnamefont {M.}~\bibnamefont {{Le
  Delliou}}}, \ and\ \bibinfo {author} {\bibfnamefont {F.~C.}\ \bibnamefont
  {Mena}},\ }\href {\doibase 10.1103/PhysRevD.88.043501} {\bibfield  {journal}
  {\bibinfo  {journal} {Phys.\ Rev.\ D}\ }\textbf {\bibinfo {volume} {88}},\
  \bibinfo {pages} {043501} (\bibinfo {year} {2013})},\ \Eprint
  {http://arxiv.org/abs/1302.6186} {arXiv:1302.6186 [gr-qc]} \BibitemShut
  {NoStop}%
\bibitem [{\citenamefont {Plebanski}\ and\ \citenamefont
  {Krasinski}(2006)}]{Plebanski:2006sd}%
  \BibitemOpen
  \bibfield  {author} {\bibinfo {author} {\bibfnamefont {J.}~\bibnamefont
  {Plebanski}}\ and\ \bibinfo {author} {\bibfnamefont {A.}~\bibnamefont
  {Krasinski}},\ }\href@noop {} {\emph {\bibinfo {title} {{An introduction to
  general relativity and cosmology}}}}\ (\bibinfo {year} {2006})\BibitemShut
  {NoStop}%
\bibitem [{\citenamefont {Maciel}\ \emph
  {et~al.}(2020{\natexlab{b}})\citenamefont {Maciel}, \citenamefont {Delliou},\
  and\ \citenamefont {Mimoso}}]{Maciel_2020}%
  \BibitemOpen
  \bibfield  {author} {\bibinfo {author} {\bibfnamefont {A.}~\bibnamefont
  {Maciel}}, \bibinfo {author} {\bibfnamefont {M.~L.}\ \bibnamefont {Delliou}},
  \ and\ \bibinfo {author} {\bibfnamefont {J.~P.}\ \bibnamefont {Mimoso}},\
  }\href {\doibase 10.1088/1361-6382/ab8759} {\bibfield  {journal} {\bibinfo
  {journal} {Classical and Quantum Gravity}\ }\textbf {\bibinfo {volume}
  {37}},\ \bibinfo {pages} {125005} (\bibinfo {year}
  {2020}{\natexlab{b}})}\BibitemShut {NoStop}%
\end{thebibliography}%
\end{document}